\documentclass[12pt]{article}
\usepackage{equations}
\usepackage[dvips]{graphicx,psfrag}
\usepackage{cite}

\begin{document}
\newlength{\captsize}           \let\captsize=\footnotesize
\newlength{\captwidth}          \setlength{\captwidth}{\textwidth}
\newlength{\beforetableskip}    \setlength{\beforetableskip}{.5\baselineskip}
\newcommand{\capt}[1]{\begin{minipage}{\captwidth}
              \let\normalsize=\captsize
              \caption[0]{#1}
              \end{minipage}\\ \vspace{\beforetableskip}}

\makeatletter
      \long\def\@makecaption#1#2{\vskip 10 \p@
      \setbox\@tempboxa\hbox{\textbf{#1:} #2}
      \ifdim \wd\@tempboxa >\hsize
            \textbf{#1:} #2\par                 
      \else
         \hbox to \hsize{\box\@tempboxa\hfil}
      \fi}
\makeatother

\renewcommand{\thefootnote}{\#\arabic{footnote}}
\setcounter{footnote}{0}

\newcommand{\slashp}{\not{\hbox{\kern-3pt $P$}}}
\newcommand{\slashs}{\not{\hbox{\kern-3pt $S$}}}

\def\a{a}
\def\tbtb{t\anti b \, \anti t b}
\def\bbbb{b\anti b b\anti b}
\def\bb{b\anti b}
\def\vtotbar{\overline V_{\rm tot}}
\def\dbar{\overline D}
\def\del{\delta}

\def\pzero{P^0}
\def\mpzero{m_{\pzero}}
\def\ntc{N_{TC}}

\def\mx{M_X}
\def\delm{\Delta m}
\def\sigmin{\sigma_{\rm min}}
\def\sigsm{\sigma_{\rm SM}}
\def\phiz{\phi_Z^0}
\def\mz{m_Z}
\def\ki{K_i}
\def\kiform{|\langle \phiz|h_i^0\rangle|^2}
\def\hi{h_i^0}
\def\mhi{m_{\hi}}
\def\hk{h_k^0}
\def\mhk{m_{\hk}}
\def\hone{h_1^0}
\def\mhone{m_{\hone}}
\def\h{h}
\def\mh{m_{\h}}
\def\gamh{\Gamma_{\h}^{\rm tot}}
\def\mhmax{\mh^{\rm max}}
\def\mhmin{\mh^{\rm min}}
\def\elep{E_{LEP}} 
\def\lam{\lambda}

\def\sqr{\wtil q_R}
\def\sql{\wtil q_L}
\def\aqp{\anti q^{\,\prime}}
\def\qp{q^\prime}
\def\qpp{q^{\prime\prime}}
\def\aqpp{\anti q^{\,\prime\prime}}
\def\asqr{\wtil q_R^*}

\def\cosa{\cos\alpha}
\def\cosb{\cos\beta}
\def\sina{\sin\alpha}
\def\sinb{\sin\beta}
\def\calb{{\cal B}}
\def\lamtil{\lam_{345}}

\def\cbpa{c_{\beta+\alpha}}
\def\cbma{c_{\beta-\alpha}}
\def\sbpa{s_{\beta+\alpha}}
\def\sbma{s_{\beta-\alpha}}
\def\ctwob{c_{2\beta}}
\def\cfourb{c_{4\beta}}
\def\stwob{s_{2\beta}}
\def\sfourb{s_{4\beta}}
\def\ctwoa{c_{2\alpha}}
\def\cfoura{c_{4\alpha}}
\def\stwoa{s_{2\alpha}}
\def\sfoura{s_{4\alpha}}
\def\ctwobma{c_{2(\beta-\alpha)}}
\def\stwobma{s_{2(\beta-\alpha)}}
\def\sb  {s_{\beta}}
\def\cb  {c_{\beta}}
\def\stwob  {s_{2\beta}}
\def\ctwob  {c_{2\beta}}
\def\sa  {s_{\alpha}}
\def\ca  {c_{\alpha}}
\def\sab  {s_{\alpha+\beta}}
\def\cab  {c_{\alpha+\beta}}
\def\sba  {s_{\beta-\alpha}}
\def\cba  {c_{\beta-\alpha}}

\def\nn{\nonumber}
\def\mup{\mu^+}
\def\mum{\mu^-}
\def\taup{\tau^+}
\def\taum{\tau^-}
\def\wpm{W^{\pm}}
\def\hpm{H^{\pm}}
\def\mhm{m_{\hm}}
\def\call{{\cal L}}
\def\calm{{\cal M}}
\def\cala{{\cal A}}
\def\wtil{\widetilde}
\def\what{\widehat}
\def\tauptaum{\tau^+\tau^-}
\def\mbb{m_{b\anti b}}

\def\bm{\boldmath}
\def\bfbm{\bf\boldmath}
\def\glsp{$\wtil g$-LSP}
\def\lsim{\mathrel{\raise.3ex\hbox{$<$\kern-.75em\lower1ex\hbox{$\sim$}}}}
\def\gsim{\mathrel{\raise.3ex\hbox{$>$\kern-.75em\lower1ex\hbox{$\sim$}}}}
\def\ifmath#1{\relax\ifmmode #1\else $#1$\fi}
\def\half{\ifmath{{\textstyle{1 \over 2}}}}
\def\threehalf{\ifmath{{\textstyle{3 \over 2}}}}
\def\quarter{\ifmath{{\textstyle{1 \over 4}}}}
\def\sixth{\ifmath{{\textstyle{1 \over 6}}}}
\def\third{\ifmath{{\textstyle{1 \over 3}}}}
\def\twothirds{{\textstyle{2 \over 3}}}
\def\fivethirds{{\textstyle{5 \over 3}}}
\def\fourth{\ifmath{{\textstyle{1\over 4}}}}

\def\gprsq{g^{\prime\,2}}
\def\dmchi{\Delta m_{\tilde\chi}}
\def\rtil{\wt r}
\def\thetaw{\theta_W}
\def\ibid{{\it ibid.}}
\def\mtil{\widetilde m}
\def\ejet{E_{\rm jet}}
\def\thetamuid{\theta(\mu\mbox{id})}
\def\mrecoil{M_{\rm recoil}}
\def\sigp{\sigma_{\rm P}^{\rm ann}}
\def\signp{\sigma_{\rm NP}^{\rm ann}}                                      
\def\alsp{\alpha_s^{\rm P}}
\def\alsnp{\alpha_s^{\rm NP}}
\def\mpi{m_{\pi}}
\def\sigann{\sigma^{\rm ann}}
\def\water{{\rm H}_2{\rm 0}}
\def\nmess{N_m}
\def\vev#1{\langle #1 \rangle}
\def\lam{\lambda}
\def\lamu{\lam_u}
\def\lamd{\lam_d}
\def\lamud{\lam_u^\dagger}
\def\lamdd{\lam_d^\dagger}
\def\lampr{\lam^\prime}
\def\lampp{\lam^{\prime\prime}}

\def\Eq#1{Eq.~(\ref{#1})}
\def\Ref#1{Ref.~\cite{#1}}

\def\bml{\hbox{$B\!\!-\!\!L$}}
\def\gtino{\wt G}
\def\mgtino{m_{\gtino}}
\def\mplanck{M_{\rm Planck}}
\def\mpl{\mplanck}
\def\sur{{\wt u_R}}
\def\msur{{m_{\sur}}}
\def\stl{{\wt t_L}}
\def\str{{\wt t_R}}
\def\mstl{m_{\stl}}
\def\mstr{m_{\str}}
\def\sbl{{\wt b_L}}
\def\sbr{{\wt b_R}}
\def\msbl{m_{\sbl}}
\def\msbr{m_{\sbr}}
\def\sq{\wt q}
\def\sqbar{\ov{\sq}}
\def\msq{m_{\sq}}
\def\slep{\wt \ell}
\def\slepbar{\ov{\slep}}
\def\mslep{m_{\slep}}
\def\slepl{\wt \ell_L}
\def\mslepl{m_{\slepl}}
\def\slepr{\wt \ell_R}
\def\mslepr{m_{\slepr}}

\def\sel{\wt e}
\def\selbar{\ov{\sel}}
\def\msel{m_{\sel}}
\def\sell{\wt e_L}
\def\msell{m_{\sell}}
\def\selr{\wt e_R}
\def\mselr{m_{\selr}}

\def\cptwo{\wt \chi^+_2}
\def\cmtwo{\wt \chi^-_2}
\def\cpmtwo{\wt \chi^{\pm}_2}
\def\mcptwo{m_{\cptwo}}
\def\mcpmtwo{m_{\cpmtwo}}
\def\stautwo{\wt \tau_2}
\def\mstautwo{m_{\stauone}}

\def\mth{m_{3/2}}
\def\delgs{\delta_{GS}}
\def\kpr{K^\prime} 

\def\caln{{\cal N}}
\def\cald{{\cal D}}
\def\DM{D$^-$}
\def\DP{D$^+$}
\def\NSM{NS$^-$}
\def\NSP{NS$^+$}
\def\HSM{HS$^-$}
\def\HSP{HS$^+$}

\def\twoloop{two-loop/RGE-improved}
\def\Twoloop{Two-loop/RGE-improved}

\def\mhi{m_{h_1^0}}
\def\etmiss{/ \hskip-8pt E_T}
\def\etmin{/ \hskip-8pt E_T^{\rm min}}
\def\etjet{E_T^{\rm jet}}
\def\ptmiss{/ \hskip-8pt p_T}
\def\mslash{/ \hskip-8pt M}
\def\rslash{/ \hskip-8pt R}
\def\susyslash{\susy\hskip-24pt/\hskip19pt}
\def\mmissl{M_{miss-\ell}}
\def\mhalf{m_{1/2}}
\def\gl{\wt g}
\def\mgl{m_{\gl}}

\def\lefft{L_{\rm eff}(t\anti t\h)}
\def\leffzh{L_{\rm eff}(Z\h)}
\def\sigmat{\sigma_T(t\anti t\h)}
\def\sigmazh{\sigma_T(Z\h)}
\def\thdm{2HDM}

\def\etc{{\em etc.}}
\def\chisq{\chi^2}
\def\cale{{\cal E}}
\def\calo{{\cal O}}
\def\eg{{\it e.g.}}
\def\etal{{\it et al.}}
\def\mhalf{m_{1/2}}
\def\dmm{\Delta^{--}}
\def\dm{\Delta^{-}}
\def\mdmm{m_{\dmm}}
\def\hdmm{h^{\dmm}}
\def\dpp{\Delta^{++}}
\def\delp{\Delta^{+}}
\def\delm{\Delta^{-}}
\def\mdelm{m_{\delm}}
\def\hzero{\Delta^0}
\def\sigdmmbar{\overline\sigma_{\dmm}}
\def\gamdmm{\Gamma_{\dmm}^T}

\def\stop{\wt t}
\def\stopone{\wt t_1}
\def\stoptwo{\wt t_2}
\def\mstop{m_{\stop}}
\def\msquark{m_{\wt q}}
\def\mstopone{m_{\stopone}}
\def\mstoptwo{m_{\stoptwo}}

\def\sbot{\wt b}
\def\sbotone{\wt b_1}
\def\sbottwo{\wt b_2}
\def\msbot{m_{\sbot}}
\def\msbotone{m_{\sbotone}}
\def\msbottwo{m_{\sbottwo}}

\def\To{\Rightarrow}
\def\lra{\leftrightarrow}
\def\msusy{m_{\rm SUSY}}
\def\msusyslash{m_{\susyslash}}
\def\susy{{\rm SUSY}}

\def\rpm{R^{\pm}}
\def\mrpm{m_{\rpm}}
\def\rzero{R^0}
\def\mrzero{m_{\rzero}}
\def\chisq{\chi^2}
\def\cale{{\cal E}}
\def\calo{{\cal O}}
\def\eg{{\it e.g.}}
\def\etal{{\it et al.}}
\def\mhalf{m_{1/2}}
\def\gl{\wt g}
\def\mgl{m_{\gl}}
\def\msquark{m_{\wt q}}
\def\mstopone{m_{\stopone}}
\def\sqbar{\ov{\sq}}
\def\slep{\wt \ell}
\def\slepbar{\ov{\slep}}
\def\mslep{m_{\slep}}

\def\hsm{h_{\rm SM}}
\def\mhsm{m_{\hsm}}
\def\hl{h^0}
\def\hh{H^0}
\def\ha{A^0}
\def\hp{H^+}
\def\hm{H^-}
\def\hpm{H^{\pm}}
\def\mhl{m_{\hl}}
\def\mhh{m_{\hh}}
\def\mha{m_{\ha}}
\def\mhp{m_{\hp}}
\def\mhpm{m_{\hpm}}
\def\tanb{\tan\beta}
\def\cotb{\cot\beta}
\def\mt{m_t}
\def\mb{m_b}
\def\mz{m_Z}
\def\mw{m_W}
\def\mgut{M_U}
\def\mx{M_X}
\def\mstring{M_S}
\def\wp{W^+}
\def\wm{W^-}
\def\wpm{W^{\pm}}
\def\wmp{W^{\mp}}
\def\chitil{\wt\chi}

\def\cnone{\wt\chi^0_1}
\def\cnonestar{\wt\chi_1^{0\star}}
\def\cntwo{\wt\chi^0_2}
\def\cnthree{\wt\chi^0_3}
\def\cnfour{\wt\chi^0_4}
\def\snu{\wt\nu}
\def\snul{\wt\nu_L}
\def\msnul{m_{\snul}}

\def\snue{\wt\nu_e}
\def\snuel{\wt\nu_{e\,L}}
\def\msnuel{m_{\snul}}

\def\snubar{\ov{\snu}}
\def\msnu{m_{\snu}}
\def\mcnone{m_{\cnone}}
\def\mcntwo{m_{\cntwo}}
\def\mcnthree{m_{\cnthree}}
\def\mcnfour{m_{\cnfour}}
\def\wt{\widetilde}
\def\wh{\widehat}
\def\cpone{\wt \chi^+_1}
\def\cmone{\wt \chi^-_1}
\def\cpmone{\wt \chi^{\pm}_1}
\def\mcpone{m_{\cpone}}
\def\mcpmone{m_{\cpmone}}

\def\staur{\wt \tau_R}
\def\staul{\wt \tau_L}
\def\stau{\wt \tau}
\def\mstaur{m_{\staur}}
\def\stauone{\wt \tau_1}
\def\mstauone{m_{\stauone}}
\def\emem{e^-e^-}
\def\sigdmmbar{\overline\sigma_{\dmm}}
\def\gamdmm{\Gamma_{\dmm}^T}

\def\MPL #1 #2 #3 {{\sl Mod.~Phys.~Lett.}~{\bf#1} (#3) #2}
\def\NPB #1 #2 #3 {{\sl Nucl.~Phys.}~{\bf #1} (#3) #2}
\def\PLB #1 #2 #3 {{\sl Phys.~Lett.}~{\bf #1} (#3) #2}
\def\PR #1 #2 #3 {{\sl Phys.~Rep.}~{\bf#1} (#3) #2}
\def\PRD #1 #2 #3 {{\sl Phys.~Rev.}~{\bf #1} (#3) #2}
\def\PRL #1 #2 #3 {{\sl Phys.~Rev.~Lett.}~{\bf#1} (#3) #2}
\def\RMP #1 #2 #3 {{\sl Rev.~Mod.~Phys.}~{\bf#1} (#3) #2}
\def\ZPC #1 #2 #3 {{\sl Z.~Phys.}~{\bf #1} (#3) #2}
\def\IJMP #1 #2 #3 {{\sl Int.~J.~Mod.~Phys.}~{\bf#1} (#3) #2}
\def\NIM #1 #2 #3 {{\sl Nucl.~Inst.~and~Meth.}~{\bf#1} {#3} #2}

\def\ltot{L_{\rm tot}}
\def\taup{\tau^+}
\def\taum{\tau^-}
\def\lam{\lambda}
\def\br{B}
\def\tauptaum{\tau^+\tau^-}
\def\sprime{{s^\prime}}
\def\rtsprime{\sqrt{\sprime}}
\def\shat{{\hat s}}
\def\rtshat{\sqrt{\shat}}
\def\gam{\gamma}
\def\sigrts{\sigma_{\tiny\rts}^{}}
\def\sigrtssq{\sigma_{\tiny\rts}^2}
\def\sigrtsprime{\sigma_{E}}
\def\nsigrts{n_{\sigrts}}
\def\betao{{\beta_0}}
\def\rhoo{{\rho_0}}
\def\etal{{\it et al.}}
\def\sighbar{\overline \sigma_{\h}}
\def\sighlbar{\overline \sigma_{\hl}}
\def\sighhbar{\overline \sigma_{\hh}}
\def\sighabar{\overline \sigma_{\ha}}
\def\anti{\overline}
\def\epem{e^+e^-}
\def\mupmum{\mu^+\mu^-}
\def\zstar{Z^\star}
\def\wstar{W^\star}
\def\zstarp{Z^{(\star)}}
\def\wstarp{W^{(\star)}}
\def\mupmum{\mu^+\mu^-}
\def\lplm{\ell^+\ell^-}
\def\brwweff{\br_{WW}^{\rm eff}}
\def\brzzeff{\br_{ZZ}^{\rm eff}}
\def\mstar{M^{\star}}
\def\mstarmin{M^{\star\,{\rm min}}}
\def\drts{\Delta\sqrt s}
\def\rts{\sqrt s}
\def\ie{{\it i.e.}}
\def\eg{{\it e.g.}}
\def\eps{\epsilon}
\def\anti{\overline}
\def\gamsnu{\Gamma_{\snu}^{\rm tot}}
\def\ai{a_1}
\def\aii{a_2}
\def\mai{m_{\ai}}
\def\maii{m_{\aii}}

\def\gamhsm{\Gamma_{\hsm}^{\rm tot}}
\def\gamhl{\Gamma_{\hl}^{\rm tot}}
\def\gamha{\Gamma_{\ha}^{\rm tot}}
\def\gamhh{\Gamma_{\hh}^{\rm tot}}

\def\fbi{{~{\rm fb}^{-1}}}
\def\fb{{~{\rm fb}}}
\def\abi{{~{\rm ab}^{-1}}}
\def\ab{{~{\rm ab}}}
\def\pbi{{~{\rm pb}^{-1}}}
\def\pb{{~{\rm pb}}}
\def\mev{{~{\rm MeV}}}
\def\gev{{~{\rm GeV}}}
\def\tev{{~{\rm TeV}}}
\def\mt{m_t}
\def\mb{m_b}

\def\hi{\h_1}
\def\hii{\h_2}
\def\hiii{\h_3}
\def\mhi{m_{\hi}}
\def\mhii{m_{\hii}}
\def\mhiii{m_{\hiii}}
\def\mpp{m_{PP}}

\def\dmm{\Delta^{--}}
\def\mdmm{m_{\dmm}}
\def\hdmm{h^{\dmm}}
\def\dpp{\Delta^{++}}
\def\mdm{m_{\dm}}
\def\hzero{\Delta^0}
\def\sigdmmbar{\overline\sigma_{\dmm}}
\def\gamdmm{\Gamma_{\dmm}^T}

\newcommand{\nc}{\newcommand}
\nc{\beq}{\begin{equation}}   \nc{\eeq}{\end{equation}}
\nc{\bea}{\begin{eqnarray}}   \nc{\eea}{\end{eqnarray}}
\nc{\baa}{\begin{array}}      \nc{\eaa}{\end{array}}
\nc{\bit}{\begin{itemize}}    \nc{\eit}{\end{itemize}}
\nc{\ben}{\begin{enumerate}}  \nc{\een}{\end{enumerate}}
\nc{\bce}{\begin{center}}     \nc{\ece}{\end{center}}
\def\beqa{\begin{eqnarray}}
\def\eeqa{\end{eqnarray}}
\def\bed{\begin{description}}
\def\eed{\end{description}}

\def\mhi{m_{h_1^0}}
\def\slash#1{#1\hskip-9pt/\hskip5pt}
\def\ptmiss{\slash p_T}
\def\etmiss{\slash E_T}
\def\mslash{\slash M}
\def\etmin{\slash E_T^{\rm min}}
\def\rslash{R\hskip-9pt/\hskip5pt}
\def\susyslash{\susy\hskip-24pt/\hskip19pt}
\def\mmissl{M_{miss-\ell}}
\def\mhalf{m_{1/2}}
\def\gl{\wt g}
\def\mgl{m_{\gl}}

\def\chisq{\chi^2}
\def\cale{{\cal E}}
\def\calo{{\cal O}}
\def\eg{{\it e.g.}}
\def\etal{{\it et al.}}
\def\mhalf{m_{1/2}}
\def\dmm{\Delta^{--}}
\def\dm{\Delta^{-}}
\def\mdmm{m_{\dmm}}
\def\hdmm{h^{\dmm}}
\def\dpp{\Delta^{++}}
\def\delp{\Delta^{+}}
\def\delm{\Delta^{-}}
\def\mdelm{m_{\delm}}
\def\hzero{\Delta^0}
\def\sigdmmbar{\overline\sigma_{\dmm}}
\def\gamdmm{\Gamma_{\dmm}^T}

\def\stl{{\wt t_L}}
\def\str{{\wt t_R}}
\def\mstl{m_{\stl}}
\def\mstr{m_{\str}}
\def\sbl{{\wt b_L}}
\def\sbr{{\wt b_R}}
\def\msbl{m_{\sbl}}
\def\msbr{m_{\sbr}}
\def\sq{\wt q}
\def\sqbar{\ov{\sq}}
\def\msq{m_{\sq}}
\def\slep{\wt \ell}
\def\slepbar{\ov{\slep}}
\def\mslep{m_{\slep}}
\def\slepl{\wt \ell_L}
\def\mslepl{m_{\slepl}}
\def\slepr{\wt \ell_R}
\def\mslepr{m_{\slepr}}

\def\stauone{\wt \tau_1}
\def\mstauone{m_{\stauone}}
\def\emem{e^-e^-}
\def\dmm{\Delta^{--}}
\def\mdmm{m_{\dmm}}
\def\hhmm{h^{\dmm}}
\def\dpp{\Delta^{++}}
\def\delp{\Delta^{+}}
\def\delm{\Delta^{-}}
\def\mdelm{m_{\delm}}
\def\hzero{\Delta^0}
\def\sigdmmbar{\overline\sigma_{\dmm}}
\def\gamdmm{\Gamma_{\dmm}^T}

\def\dmm{\Delta^{--}}
\def\dmml{\Delta^{--}_L}
\def\mdmm{m_{\dmm}}
\def\mdmml{m_{\dmml}}
\def\hdmm{h^{\dmm}}
\def\dpp{\Delta^{++}}
\def\dppl{\dpp_L}
\def\mdm{m_{\dm}}
\def\hzero{\Delta^0}
\def\hzerol{\Delta^0_L}
\def\sigdmmbar{\overline\sigma_{\dmm}}
\def\gamdmm{\Gamma_{\dmm}^T}
\def\gamdmml{\Gamma_{\dmml}^T}
\def\mm{\mu^+\mu^-}
\def\ee{e^+e^-}
\def\rta{\rightarrow}
\def\tanb{\tan\beta}

\def\mz{m_Z}
\def\mw{m_W}
\def\wp{W^+}
\def\wm{W^-}
\def\gam{\gamma}
\def\h{h}
\def\mh{m_{h}}
\def\mphi{m_\phi}
\def\what{\widehat}
\def\lwh{\widehat\Lambda_W}
\def\lphihat{\widehat\Lambda_\phi}
\def\lphi{\Lambda_\phi}
\def\mphisq{m_\phi^2}
\def\mphi{m_\phi}
\def\mbar{\overline m}
\def\hbar{\overline h}
\def\lam{\lambda}
\def\mpl{M_{Pl}}
\def\ifmath#1{\relax\ifmmode #1\else $#1$\fi}
\def\half{\ifmath{{\textstyle{1 \over 2}}}}
\def\threehalf{\ifmath{{\textstyle{3 \over 2}}}}
\def\quarter{\ifmath{{\textstyle{1 \over 4}}}}
\def\sixth{\ifmath{{\textstyle{1 \over 6}}}}
\def\third{\ifmath{{\textstyle{1 \over 3}}}}
\def\twothirds{{\textstyle{2 \over 3}}}
\def\fivethirds{{\textstyle{5 \over 3}}}
\def\fourth{\ifmath{{\textstyle{1\over 4}}}}
\def\call{{\cal L}}
\def\calo{{\cal O}}
\def\sig{\sigma}
\def\eps{\epsilon}

\def\GeV{{\rm GeV}}\def\ra   {\rightarrow}
\def\TeV{{\rm TeV}}\def\rmax{{\rm max}}\def\tq{\widetilde q}
\def \fbi{{\rm fb^{-1}}} \def\Cgrav{C_{\rm grav}}
\def\simle{
    \mathrel{\rlap{\raise 0.511ex 
        \hbox{$<$}}{\lower 0.511ex \hbox{$\sim$}}}}
\def \pbi{{\rm pb^{-1}}} 
\def\ol   {\overline}
\def\lsp{\widetilde\chi_1^0}
\def\tell{{\widetilde\ell}}
\def\tchi{\widetilde\chi}
\def\GeV{{\rm GeV}}
\def\tG{{\widetilde G}}
\def\Meff{M_{\rm eff}}
\def\sgn{\mathop{\rm sgn}}
\def\tg{\widetilde g}
\def\lsp{\widetilde\chi_1^0}
\def\slashchar#1{\setbox0=\hbox{$#1$}           
   \dimen0=\wd0                                 
   \setbox1=\hbox{/} \dimen1=\wd1               
   \ifdim\dimen0>\dimen1                        
      \rlap{\hbox to \dimen0{\hfil/\hfil}}      
      #1                                        
   \else                                        
      \rlap{\hbox to \dimen1{\hfil$#1$\hfil}}   
      /                                         
   \fi}  
\def  \abseta {\mid\eta\mid} \def\tchi{\widetilde\chi}
\def\tell{{\widetilde\ell}}
\def\etmiss{\slashchar{E}_T}

\renewcommand{\thefootnote}{\fnsymbol{footnote}}
\setcounter{footnote}{0}

\def\thefootnote{\fnsymbol{footnote}}

%
%

\vskip 1. truecm
\begin{flushright}
\noindent UCD-02-19 \par
\noindent December, 2002 \par
\end{flushright}

\vskip 1in
\begin{center}
{\Large \bf Extended Higgs Sectors}~\footnote{To appear in the Proceedings
of {\it SUSY02: The 10th International Conference on Supersymmetry and Unification of Fundamental Interactions}, June 17-23, 2002, DESY Hamburg.}\\[1cm]
{\large John F. Gunion}\\[5pt]
{\it Department of Physics, University of California at Davis, Davis, CA 95616, USA}\\
\end{center}

\begin{abstract}
I give a brief overview of the motivations for and experimental probes
of extended Higgs sectors containing more than the single Higgs
doublet field of the Standard Model.
\end{abstract}

We are now approaching the 40th anniversary of the introduction~\cite{higgs,eb}
of the idea of a Higgs mechanism for electroweak symmetry breaking
and mass generation  using an elementary scalar field~\cite{hhg}.
Remarkably, we still have no experimental information that either
confirms or excludes the elementary scalar Higgs boson(s) that 
would arise if this mechanism is correct. All current experimental
data is consistent with the Standard Model (SM) and its single
CP-even Higgs boson ($\hsm$) provided $\mhsm\lsim 200\gev$,
for consistency with precision electroweak analyzes~\cite{precision},
and $\mhsm\gsim 114\gev$, for consistency with direct limits
from LEP2~\cite{LEPHiggs}. The Tevatron and LHC will be probing
the remaining allowed $\mhsm$ mass region over the next decade.
However, there are many theoretical reasons supporting the idea that
the SM with its simple one-doublet Higgs sector is not the whole story.
In particular, the SM has difficulty explaining a light $\hsm$ (the
naturalness and hierarchy problems) and does not predict gauge
coupling unification.
Thus, theorists have spent many years constructing extensions
of the SM that rectify these two inadequacies. Most involve
an extension of the Higgs sector to include at least two Higgs
doublets, and perhaps singlets and/or triplet and other representations.
Here, I will survey some of the ideas and associated experimental challenges.
The most important conclusion will be that only a combination of
the LHC and a future linear collider (LC) is guaranteed to
find a Higgs boson and that full exploration of the Higgs sector
might require both machines plus a $\gam\gam$ collider facility ($\gam C$)
at the LC, and possibly a muon collider ($\mu C$).

We begin with a discussion of whether or not we need supersymmetry
for gauge coupling unification and a solution to the fine-tuning problem.
This will lead to some specific topics regarding
complicated Higgs sectors and extra-dimensional theories.
We then consider what it will take to fully explore the Higgs sector
of the minimal supersymmetric model (MSSM), the scalar sector for which is
a highly constrained two-Higgs-doublet model (2HDM). Next, we review the
strongly motivated next-to-minimal supersymmetric model (NMSSM) 
in which one singlet superfield is added to the MSSM,
and the experimental consequences of the extra Higgs bosons that
arise. The fully general case of multiple Higgs singlets, as often
found in string-motivated models, is then considered.
We end with a reminder regarding the somewhat overlooked, but interesting,
left-right supersymmetric models that allow for automatic
solutions of several important problems. 

\vspace*{-.1in}
\section{\bfbm Extensions of the SM Higgs sector}

In this section, we discuss various models 
and their motivations in which the only
extension of the SM is the addition of more Higgs representations
to the Higgs sector.

\smallskip
\noindent{\it Motivations from coupling constant unification}
\smallskip

Coupling constant unification can be achieved
simply by introducing additional Higgs representations in the 
SM~\cite{Gunion:1995mq,Gunion:1998ii}.
For $\rho=1$ to be natural, the neutral field member (if there is one)
of representations other than $T=1/2,|Y|=1$ should have zero vacuum expectation
value (vev)~\cite{Gunion:1991dt}. Some simple choices for representations 
that yield coupling unification are shown
in Table~\ref{repu}. There, $N_{T,Y}$ gives the number of representations
with the indicated weak-isospin $T$ and hypercharge $Y$.
From the table, we observe that achieving coupling constant unification
in this way requires a lower unification scale, $M_U$, 
than comfortable for proton decay. This need
not be a problem if the coupling unification is not associated
with true group unification (\ie\ if there are no extra $X,Y$ gauge
bosons to mediate proton decay), as is possible in some string theory models.
The solution with the largest $M_U$ is $N_{1/2,1}=2$ and $N_{1,0}=1$.
Many of the most attractive solutions contain several doublets
and one or more triplets.
With sufficiently complicated Higgs sectors, we can even achieve
coupling unification at very low $M_U$ values, as possibly 
appropriate in large-scale extra-dimension models.
Another motivation for models with two or more Higgs doublets is that
both explicit and spontaneous CP violation in the Higgs sector
is possible (see, for example, \cite{hhg,Grzadkowski:1999ye,Gunion:2002zf}).
Of course, once one has two or more doublets in the Higgs sector,
there will be many Higgs potential parameters.  These must
be constrained so that the potential minimum is such that
all Higgs bosons have positive mass-squared.
In particular, $\mhpm^2>0$ is required in order to avoid
breaking of electromagnetism.

\begin{table}[t!]
\capt{\label{repu} Choices of Higgs representations that
allow for coupling constant unification without any other extension
of the SM. $N_{T,Y}$ is the number of representations with indicated $T$
and $Y$. The tabulated $\alpha_s(\mz)$ values are those that
allow for unification at the tabulated $\mgut$ scales.}
\begin{center}
\begin{tabular}{|c|c|c|c|c|c|c|c|}
\hline
$N_{1/2,1}$ & $N_{1/2,3}$ & $N_{0,2}$ & $N_{0,4}$ & $N_{1,0}$ & 
$N_{1,2}$ & $\alpha_s(\mz)$ & $\mgut$ (GeV) \cr
\hline
1 & 0 & 0 & 2 & 0 & 0 & 0.106 & $4\times 10^{12}$ \cr
1 & 0 & 4 & 0 & 0 & 1 & 0.112 & $7.7\times 10^{12}$ \cr
1 & 0 & 0 & 0 & 0 & 2 & 0.120 & $1.6\times 10^{13}$ \cr
2 & 0 & 0 & 0 & 1 & 0 & 0.116 & $1.7\times 10^{14}$ \cr
2 & 0 & 2 & 0 & 0 & 2 & 0.116 & $4.9\times 10^{12}$ \cr
2 & 1 & 0 & 0 & 0 & 2 & 0.112 & $1.7\times 10^{12}$ \cr
3 & 0 & 0 & 0 & 0 & 1 & 0.105 & $1.2\times 10^{13}$ \cr
\hline
\end{tabular}
\end{center}
\end{table}

\begin{figure}[b!]
\vspace*{-.3in}
\begin{center}
\includegraphics[width=14cm]{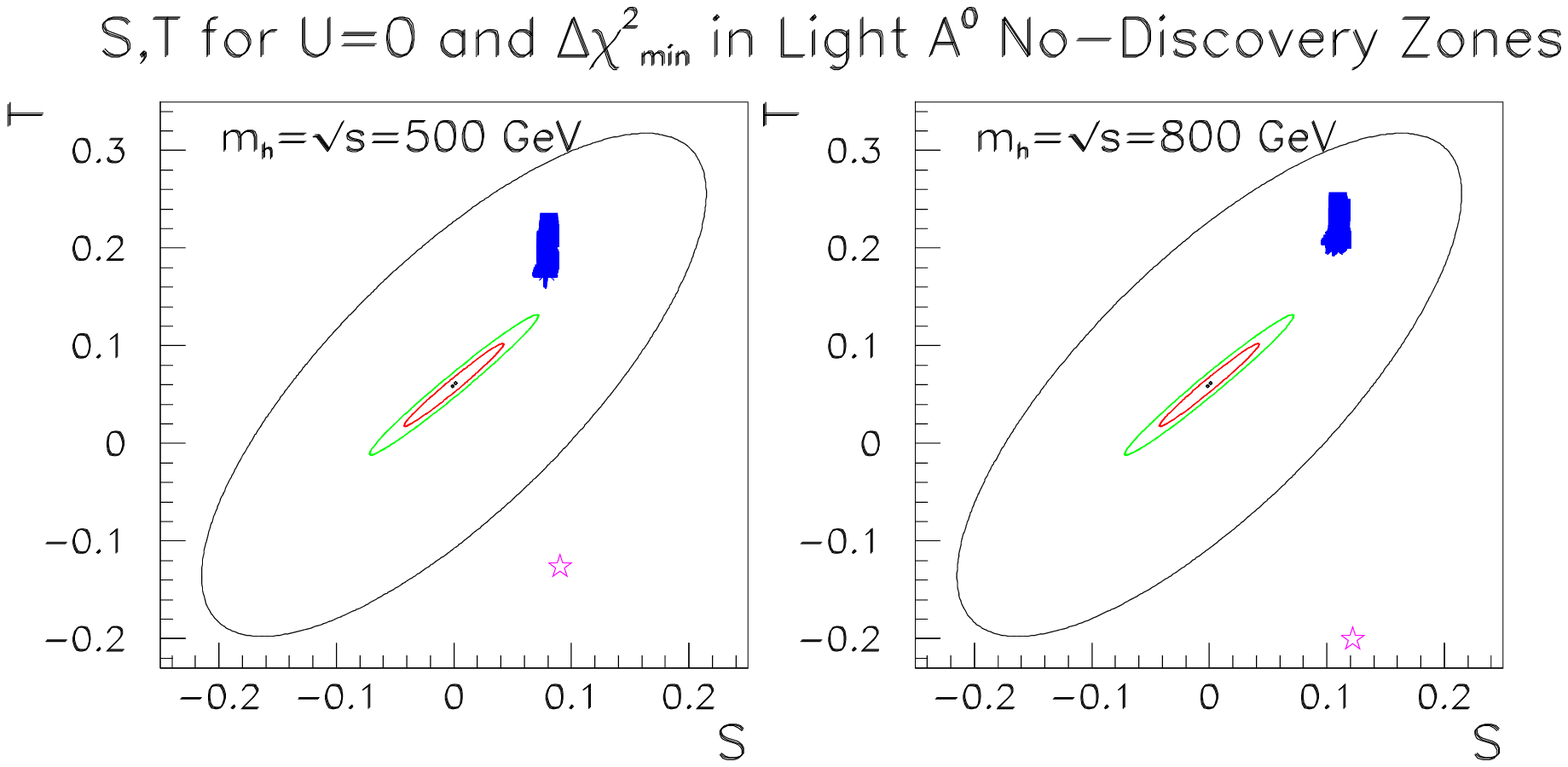}
\end{center}
\vspace*{-2.8in}
\capt{\label{ellipse}
The outer ellipses show the 90\% CL region from current
precision electroweak data in the $S,T$ plane
for $U=0$ relative to a central point defined
by the SM prediction with $\mhsm=115$ GeV. 
The blobs of points show the $S,T$ predictions for 2HDM  models 
with a light $\ha$ and with
$\tanb$ such that the $\ha$ cannot be detected
in $b\anti b\ha$ or $t\anti t\ha$ production
at either the LC or the LHC; the mass of
the SM-like $\hl$ is set equal to $\rts=500\gev$ (left)
or $800\gev$ (right) and $\mhpm$ and $\mhh$ have been
chosen to minimize the $\chi^2$ of the full precision electroweak fit.
The innermost (middle) ellipse shows the  90\% (99.9\%) CL
region for $\mhsm=115$ GeV after
Giga-$Z$  LC operation {\it and} a $\Delta m_W\protect\lsim 6$ MeV threshold
scan measurement. The stars to the bottom right show the $S,T$ predictions
in the case of the SM with $\mhsm=500\gev$ (left) or $800\gev$ (right).
This figure is from \cite{Gunion:2001vi}.} 
\end{figure}

\smallskip
\noindent{\it The light CP-odd Higgs boson scenario in a general two-Higgs-doublet model}
\smallskip

Even the simple CP-conserving 2HDM extension of the SM Higgs sector allows
for some unusual scenarios.  In particular, suppose that
the $\ha$ of the 2HDM is moderately light and all other Higgs bosons are heavy.
Remarkably, this type of scenario 
can be consistent with precision electroweak
constraints~\cite{Chankowski:2000an}.
If $\mha$ is small, the best fit to the precision
electroweak data is achieved by choosing 
the lighter CP-even Higgs boson, $\hl$,
to be SM-like. A good fit is possible even
for $\mhl\sim 1\tev$. Of course, 
such a heavy SM-like $\hl$ leads to large 
$\Delta S>0$ and large $\Delta T<0$ contributions, which 
on their own would place the $S,T$ prediction of the 2HDM model 
well outside the current 90\% CL ellipse --- see the stars in 
Fig.~\ref{ellipse}, taken from \cite{Gunion:2001vi}.
However, the large $\Delta T<0$ contribution from the SM-like $\hl$
can be compensated by a large $\Delta T>0$ from a
small mass non-degeneracy (weak isospin breaking) of the
still heavier $\hh$ and $\hpm$ Higgs bosons. In detail, for a 
moderate light $\ha$ (roughly $\mha\lsim \half\mhl$)
and SM-like $\hl$ one finds
\beq
   \Delta \rho=\frac{\alpha}{16 \pi m_W^2 c_W^2}\left\{\frac{c_W^2}{s_W^2}
   \frac{m_{H^\pm}^2-m_{H^0}^2}{2}-3m_W^2\left[\log\frac{m_{h^0}^2}{m_W^2}
   +\frac{1}{6}+\frac{1}{s_W^2}\log\frac{m_W^2}{m_Z^2}\right]\right\}\nonumber
\label{drhonew}
\eeq
from which we see that the first term can easily compensate
the large negative contribution to $\Delta\rho$ from the $\log (\mhl^2/\mw^2)$
term. In Fig.~\ref{ellipse}, 
the blobs correspond to 2HDM parameter choices for which:
(a) $\mhl=\rts$ (either $500\gev$ or $800\gev$)
of a linear $\epem$ collider (\ie\ $\mhl$
is such that the $\hl$ cannot be observed at the LC); (b)
$\mhpm-\mhh\sim {\rm few}\gev$ has been chosen
(with both $\mhpm,\mhh\gsim 1\tev$)
so that the $S,T$ prediction is well within the 90\% CL ellipse
of the precision electroweak fits; and (c)
$\mha$ and $\tanb$ are in the `wedge' of $[\mha,\tanb]$ parameter space
characterized by moderate $\tanb$ values and $\mha\gsim 250\gev$
for which the LHC and $\epem$ LC operation at $\rts=500\gev$ or $800\gev$
would not allow discovery of the $\ha$ through $b\anti b\ha$
or $t\anti t\ha$ production \cite{Grzadkowski:2000wj} (see also
~\cite{Djouadi:gp}) and the LC $\epem\to Z\ha\ha$ and $\epem\to \nu\anti\nu
\ha\ha$ rates are too small to be detected (as is the case
for $\mha\gsim 150\gev$ 
at $\rts=500\gev$ and $\mha\gsim 270\gev$ at $\rts=800\gev$)~\cite{Haber:1993jr,Djouadi:1996jf,Farris:2002ny}.
However, this scenario
can only be pushed so far. In order to maintain perturbativity
for all the Higgs self couplings, it is necessary that the 
quartic couplings of the 2HDM potential obey 
$|\lam_i|/(4\pi)\lsim \mathcal{O}(1)$~\cite{weldon,arhrib,Kanemura:1993hm,Gunion:2002zf}. 
This in turn implies that the $\hl$, $\hh$
and $\hpm$ masses should obey $\mhl,\mhh,\mhpm\sim
|\lam_i|^{1/2} v\lsim 800-900\gev$. This bound on $\mhl$
also ensures the absence of strong $WW$ scattering --- see \cite{hhg}.
Thus, the SM-like $\hl$ would be detected at the LHC.
If it should happen that a heavy SM-like Higgs boson is detected
at the LHC, but no other new particles (supersymmetric particles,
additional Higgs bosons, \etc) are observed, 
the precision electroweak situation could 
only be resolved by Giga-$Z$ operation 
and a $\Delta\mw=6\mev$ $WW$ threshold scan at the LC
(yielding the 90\% CL Giga-$Z$ 
ellipse sizes illustrated in Fig.~\ref{ellipse}).
The resulting determination of $S,T$ 
would be sufficiently precise to definitively check for values
like those of the blobs of Fig.~\ref{ellipse}. If no
other new physics was detected at the LC or LHC that could cause
the extra $\Delta T>0$, searching for the other Higgs
bosons of a possible 2HDM Higgs sector, 
especially a relatively light decoupled $\ha$, would become a high priority. 
Interestingly, the  current discrepancy 
with SM predictions for  $a_\mu$ can be explained 
in whole or part~\footnote{For the rather 
low $\mha$ and high $\tanb$ values required in
order that the $\ha$ be the {\it full} explanation of the discrepancy,
the $\ha$ would be seen at the LC in $\epem\to Z\ha\ha$ 
and $\epem\to b\anti b \ha$ production
if not earlier at the LHC.} by two-loop diagrams involving 
a light $\ha$ \cite{Cheung:2001hz,Krawczyk:2001nw}.

\smallskip
\noindent{\it Special cases in which Higgs discovery would be complicated
and/or difficult}
\smallskip

Some additional complications that would make Higgs discovery
more difficult in the case of the general 2HDM or a still more 
extended Higgs sector are: 
\bed
\item{(i)} The Higgs sector could be CP-violating.

Both spontaneous and explicit CP-violation 
is possible for a general 2HDM 
(see, {\it e.g.} \cite{hhg,Grzadkowski:1999ye,Gunion:2002zf}). 
If CP-violation is present,
the three neutral Higgs bosons mix to form three mass eigenstates
of mixed CP nature, $h_{1,2,3}$, which
share the $WW/ZZ$ coupling strength squared: $\sum_i g_{VVh_i}^2=g_{VV\hsm}^2$.
In this case, the signal for any one of them would be weakened,
perhaps dramatically so.  Such sharing would be particularly devastating
for the LHC $gg\to h_i \to \gam\gam$ signals. While this would reduce
the LC $\epem\to Zh_i$ signals, the above sum rule
and the fact that the $h_i$ with large $g_{VVh_i}^2$ would need
to be light ($\lsim 200\gev$) in order to agree with precision electroweak data
(modulo the type of special situation described in the previous subsection)
imply that at least one of the signals would always be visible.

\item{(ii)} The (possibly mixed) Higgs bosons
could be sufficiently close in mass that their resonance peaks,
which have finite (decay-channel-dependent) 
width because of experimental resolution, would overlap.

At the LC, this would smear out the $\epem\to Z h_i$ signals.
As discussed later, the Higgs signal would still be revealed
as a broad enhancement (from the composite of the overlapping
signals) in the $\mx$ spectrum observed in $\epem \to Z X$ events.
At the LHC, if the $gg\to h_i \to \gam\gam$ 
signal for one or more of the $h_i$ is of observable strength,
the excellent experimental $m_{\gam\gam}$ resolution would make it
likely that each individual signal could be seen. However, this
would not be the case for $h_i\to \tau^+\tau^-$ and $b\anti b$ discovery
channels where the experimental resolution is not very good.
The related Higgs signals would be very difficult to extract.

\item{(iii)} 
All the Higgs bosons with substantial $g_{VVh_i}^2$ 
could decay to a pair of lighter Higgs bosons or to a light
Higgs boson and a gauge boson. For example, in the
CP-conserving case, we could have large $\hl,\hh\to\ha\ha$ branching
ratios. At the LHC, 
this would greatly weaken the $gg\to\hl,\hh\to \gam\gam$ signals,
which then might not be detectable. The $WW\to\hl,\hh\to \tau^+\tau^-,\bb$
signals would also be very weak. Searches for the CP-even
Higgs bosons would have to rely on the
 $\ha\ha\to\tau^+\tau^-\tau^+\tau^-$ and $\tau^+\tau^-\bb$
final states, which have not been shown to lead to observable
signals at the LHC. Existing LHC studies~\cite{LHCcms,LHCatlas} suggest that 
single $\ha$ detection in the $\tau^+\tau^-$ or $\bb$
final states would be very difficult.
In contrast, the $\hl$ or $\hh$ would be detectable
at the LC using the $\epem\to ZX$ search for a resonant bump
in $\mx$. Once found, the $\ha\ha$ decays of the $\hl$ and $\hh$ could
be studied. 

\eed
Finally, there is nothing to rule out a combination of the above
difficulties.  In such a case, only the LC would have the ability
to detect at least one of the Higgs bosons of the general 2HDM.
Still more complicated Higgs sectors,
for example one containing many doublets and a number of singlets,
would lead to still greater difficulties.
A common theme in all the above scenarios, and in the preceding light-$\ha$
2HDM scenario, is the probable importance of directly detecting a light $\ha$.
As outlined later, the $\gam C$ and $\mu C$ are likely to be the
machines of choice for this purpose unless $\tanb$ is very large.
More detailed examples
of these general complexities/difficulties will be outlined
in the supersymmetric Higgs portion of this talk.

\smallskip
\noindent{\it Triplet Higgs representations}
\smallskip

SU(2)$_L$ triplet Higgs representations ($\Delta_L$)
with zero vev for their neutral members
have significant motivation. They are
especially well-motivated in the context of
left-right (LR) symmetric and related models (see \cite{hhg}
for discussion and references).
The $2\times 2$ notation for a $T=1,|Y|=2$ triplet is 
$\Delta=\left(\begin{array}{cc} \delp/\sqrt{2} & \dpp \cr \hzero & -\delp/\sqrt{2} \cr \end{array}\right)\,.$ 
The non-supersymmetric LR model contains both $\Delta_R$ 
and $\Delta_L$ triplets. A Majorana lepton-number-violating coupling
is introduced for the $\Delta_R$ so that a large
$\vev{\Delta^0_R}$ will yield a
large Majorana $\nu_R$ mass as well as large $m_{W_R}$. 
The LR symmetry requires that the $\Delta_L$ triplet have an
equivalent lepton-number-violating coupling. However,
symmetry breaking
can be arranged so that $\vev{\Delta^0_L}=0$ (to keep $\rho=1$ natural).
More generally, there is no reason not to consider the possibility of 
a $\Delta_L$ with $\vev{\Delta^0_L}=0$. Coupling constant unification
can be arranged in models with triplets, but might require other
types of matter fields at intermediate scales and/or extra dimensional
physics~\cite{Lindner:1996tf,Perez-Lorenzana:1998rj,Perez-Lorenzana:1999qb}.

The interesting phenomenology of triplets is illustrated
by focusing on the case of a $|Y|=2$ triplet representation, for which
the lepton-number-violating coupling Lagrangian is:
\begin{equation}
{\cal L}_Y=ih_{ij}\psi^T_{i} C\tau_2\Delta\psi_{j}+{\rm h.c.}
\,,\quad i,j=e,\mu,\tau\,.
\label{couplingdef}
\end{equation}
which leads to lepton-number-violating
$\emem,\mu^-\mu^-,\tau^-\tau^-\rta\dmm$ couplings. If we write 
$|\hdmm_{\ell\ell}|^2\equiv c_{\ell\ell} \mdmm^2(\gev)\,,$ the strongest
limits are $c_{ee}<10^{-5}$ (from Bhabha scattering) and $c_{\mu\mu}< 10^{-6}$ 
[noting that a triplet gives the wrong sign for the observed
$(g-2)_\mu$ deviation]. 
For $\vev{\Delta^0_L}=0$, $\gamdmml$  would be small
and $\dmml\to \ell^-\ell^-$ decays could dominate.  
 For $\mdmml\lsim 1\tev$, one would discover the $\dmml$ in
$p p\to \dmml\dppl$ with $\dmml \to
\ell^-\ell^-,\dppl\to\ell^+\ell^+$ ($\ell=e,\mu,\tau$) at
the LHC (or earlier at the Tevatron if 
$\mdmml\lsim 350\gev$)~\cite{Gunion:1996pq}.
Thus, the $pp$ colliders will tell us if such a $\dmml$ exists in the mass
range accessible to a LC or possible $\mu C$ {\it and how it decays}.
However, only the relative $c_{\ell\ell}$ values for those
$\ell$'s observed in $\dmml\to\ell^-\ell^-$ decays could be determined.
The next step would be to
produce and study the $\dmml$ in $\ell^-\ell^-$ 
$s$-channel collisions. If $c_{ee}$ ($c_{\mu\mu}$) is near its
current upper limit, event rates in $e^-e^-$ ($\mu^-\mu^-$) collisions would 
be enormous~\cite{Gunion:1995mq,Gunion:1998ii}
for the expected small values of $\gamdmml$ and would provide a direct
measure of the corresponding $c_{\ell\ell}$.
Since backgrounds are very small,
observable signals would be present for even very small $c_{\ell\ell}$ ---
\eg\ a $c_{ee}$ value as small as $\sim 10^{-16}$ 
could be probed at an $\emem$ collider with $L=300~\fbi$.
This would cover essentially the entire range of coupling
relevant for the see-saw mechanism.

\vspace*{-.1in}
\section{\bfbm Higgs-radion mixing in the Randall-Sundrum model}

Although models in which only the Higgs sector of the SM
is extended lead to interesting new phenomenological possibilities,
they do not solve the hierarchy problem ---
there is no natural reason for Higgs boson masses to be below $\sim 1\tev$.
Large-scale extra dimensions appear to be required in order to
solve the hierarchy problem without the introduction
of supersymmetry. One model of this type is the Randall-Sundrum (RS)
model \cite{Randall:1999ee}, 
wherein a single extra (5th) dimension is introduced with a warped
metric between two 3-branes (\ie\ branes with 3 spatial dimensions
and 1 time dimension). 
In the simplest version, all SM fields are confined to the ``visible'' brane;
only gravity propagates in the 5th dimension. The TeV scale on the
visible brane arises as an exponential suppression warp factor
times the Planck scale on the ``invisible'' brane. The RS approach gives
rises to many fascinating new phenomena. Of particular interest
are the possibly dramatic implications of such a model for the Higgs sector.
If all matter (in particular the one Higgs doublet of the SM) 
is on the TeV brane, the most
interesting deviations from SM Higgs physics arise if there is mixing
of the Higgs doublet with the radion~\cite{wellsmix,csakimix,Han:2001xs,Chaichian:2001rq,Hewett:2002nk,Csaki:1999mp,Dominici:2002np,Dominici:2002jv}. The mixing arises
from the allowed action form:
\def\Hhat{\what H}
\beq
S_\xi=\xi \int d^4 x \sqrt{g_{\rm vis}}R(g_{\rm vis})\Hhat^\dagger \Hhat\,.
\eeq
Here, $R(g_{\rm vis})$ is the Ricci scalar for the metric induced 
on the visible brane, $\Hhat$ is the Higgs field (before rescaling
to canonical normalization on the brane), and 
$g^{\mu\nu}_{\rm vis}=\Omega_b^2(x)(\eta^{\mu\nu}+\eps h^{\mu\nu})$,
where the quantum fluctuations in $\Omega_b(x)$ define the radion
field (before rescaling) and the $h^{\mu\nu}$
are the fluctuations about the locally flat 4-d metric.
For $\xi\neq 0$, the Higgs and radion mix and one must rediagonalize
and rescale to canonically-normalized mass eigenstates $\h$ and $\phi$.

\begin{figure}[t!]
\begin{center}
\includegraphics[width=2.3in,height=5in,angle=90]{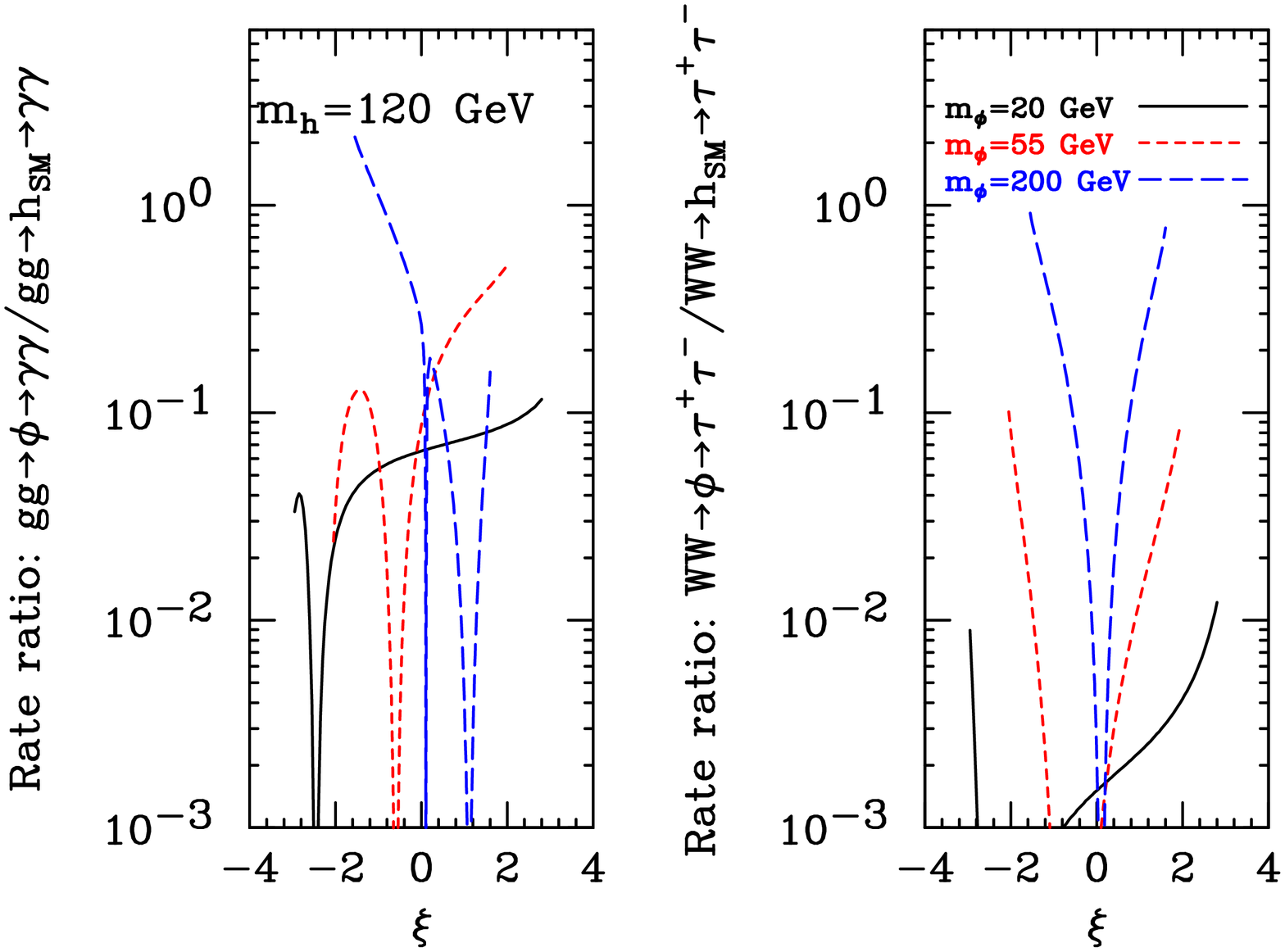}
\end{center}
\begin{center}
\includegraphics[width=2.3in,height=5in,angle=90]{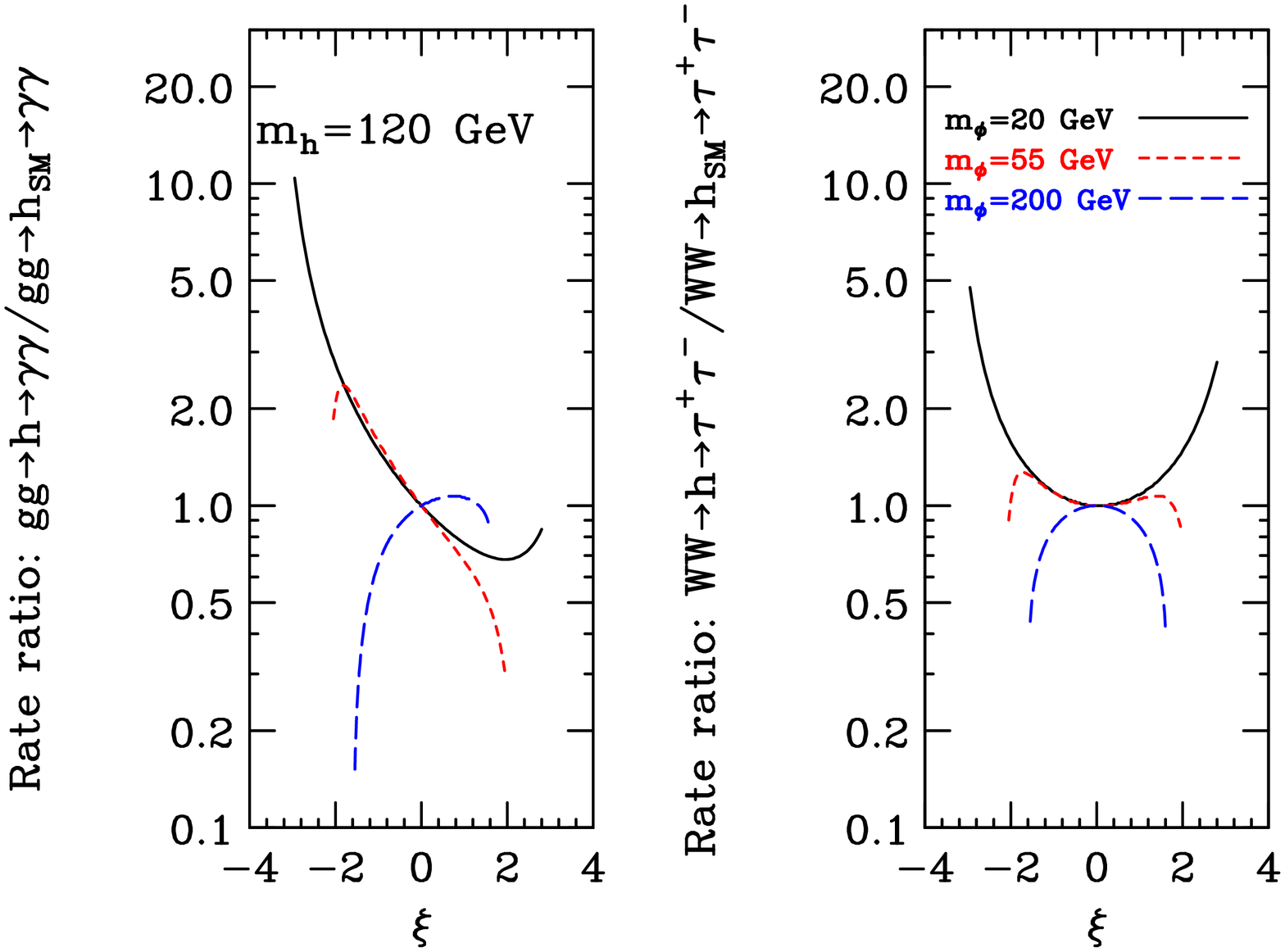}
\end{center}
\label{prodh_mh120}
\capt{\label{rsprod} The ratio of important LHC $\h$ and $\phi$ 
production/decay discovery channels to the prediction for the $\hsm$
as a function of the mixing parameter $\xi$, assuming $\mh=120\gev$
and for $\mphi=20,55,200\gev$. The $\h$ ($\phi$)
comparisons assume $\mhsm=\mh$ ($\mhsm=\mphi$), respectively.
The upper and lower
limit for $\xi$ of each curve is determined by theoretical constraints
within the RS model. From \cite{Dominici:2002np}.}
\vspace*{-.3in}
\end{figure}

The basic parameters determining the Higgs-radion
phenomenology are $\mh$, $\mphi$, $\lphi$ (the new physics
scale characterizing the radion interactions) and $\xi$.
A complicated inversion process relates these
to the bare parameters of the Lagrangian needed to compute the couplings
of the $\h$ and $\phi$.  We very briefly outline the consequences
of $\xi\neq 0$ as obtained in \cite{Dominici:2002np}.
While it is possible to have $\mh\sim 112\gev$ (\ie\ somewhat below
the SM lower limit of $114\gev$) without violating
LEP constraints on $g_{ZZ\h}^2$, let us focus on the case of $\mh=120\gev$. 
The $\h$ and $\phi$ will typically be detected in the same modes
as have been studied for the SM Higgs boson. 
For allowed $\xi$ values, the $\h$ and $\phi$ discovery mode rates
at the LHC and at the LC can be dramatically different 
as compared to the rates predicted for a $\hsm$ of the same mass.
This is illustrated in Fig.~\ref{rsprod}. This figure shows
that for most values of $\xi$ the $\phi$ rates will be much
smaller than expected for the $\hsm$ (when $\mhsm=\mphi$). However,
for some values of $\xi$ the LHC rates for the $\phi$
are closer to being SM-like than those of the $\h$. 
Typically, a LC will be required to fully unravel what is going on.
Where rates in the plotted LHC discovery modes are small for the $\phi$,
it could be that the LHC would still be able to discover the
$\phi$ in $\h\to\phi\phi\to \bb\bb,\bb gg$ decays.
The decay $\h\to\phi\phi$ can have a sizable branching
ratio and would provide a definitive signature that
$\xi\neq 0$ mixing is present. If this and other LHC signatures
for the $\phi$ are too weak to be detected,
an LC will be needed to discover the $\phi$. Indeed,
an LC with $L=500~\fbi$ can detect $\epem\to Z^*\to Z\phi$ events for
very small $g_{ZZ\phi}^2$ values, the precise limit
depending upon $\mphi$. The only part
of parameter space for which the LC could not detect the $\phi$
is in the vicinity of a line in $(\xi,\lphi)$ parameter space 
where $g_{ZZ\phi}^2=g_{f\anti f\phi}^2=0$.
This illustrates the importance of the LC to a full exploration
of the RS Higgs-radion sector. For many choices of parameters,
a $\gam C$ would be extremely valuable for sorting out the
Higgs-radion sector \cite{Asner:2002aa}. This is because one of
the most characteristic features of the RS model is
the presence of anomalous $\h\to\gam\gam,gg$ and $\phi\to\gam\gam,gg$
couplings that can only be extracted in a model-independent manner
using $gg\to\h,\phi\to\gam\gam$ and $\gam\gam\to \h,\phi\to b\anti b$
measurements.

The RS model does have some undesirable features.
In particular, there is the 
new fine-tuning problem of adjusting cosmological
constants on the branes and in the bulk to have exactly the
right relationships. A more fundamental source for
these relationships has yet to be demonstrated.
Coupling unification is also problematical in that
the couplings would only appear to unify (via logarithmic
running) at the 4-d Planck scale
or typical GUT scale if there is 
matter off the brane~\cite{Randall:2001gc,Agashe:2002bx,Goldberger:2002pc}.

\vspace*{-.1in}
\section{\bfbm Higgs sectors in supersymmetric models}

Supersymmetry is still viewed
as the best approach to solving the hierarchy and naturalness problems
and no other model yields coupling unification, and also
electroweak symmetry breaking, in such an inherently
natural way. Thus, the balance of the talk will focus on
how well we can explore a supersymmetric model Higgs sector.
We will begin with the MSSM and then move to the NMSSM and
to LR-symmetric supersymmetric models.

\smallskip
\noindent{\it  MSSM Higgs sector highlights}
\smallskip

\begin{figure}[b!]
\begin{center}
\includegraphics[width=4in]{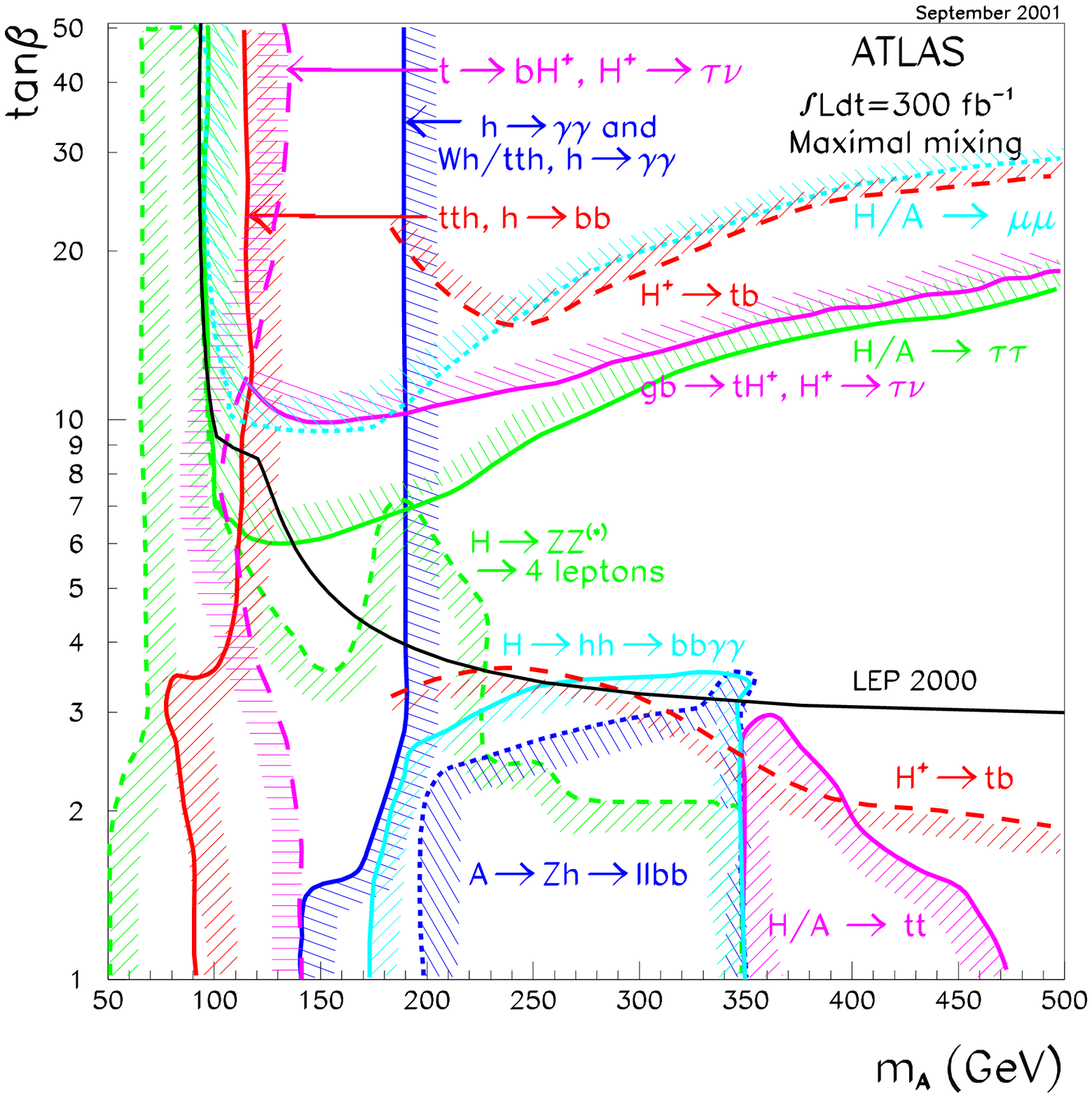}
\capt{\label{atlasmaxmix}
$5\sigma$ discovery contours for MSSM Higgs boson detection
in various channels are shown in the $[\mha,\tanb]$ parameter plane, 
assuming maximal mixing and $L=300~\fbi$
for the ATLAS detector~\cite{atlasmaxmixref}.
}
\vspace*{-.2in}
\end{center}
\end{figure}

In the case of the MSSM Higgs sector (as reviewed, for example,
in \cite{hhg,Gunion:1996cn,higgsreview,ghvk}), 
the key issue is the extent to which
we will be able to completely explore the Higgs sector
at the Tevatron, LHC and future LC. The discussion here will
assume the maximal-mixing scenario with a SUSY scale of 1 TeV,
and the absence of CP violation in the Higgs sector. In this case, 
the light CP-even $\hl$ has mass $\mhl<135\gev$.
Assuming that the CP-odd Higgs boson has mass $\mha\gsim 200\gev$
(as is probable, given typical renormalization group evolution
scenarios for electroweak symmetry breaking),
the Higgs sector will be in the decoupling regime~\cite{Gunion:2002zf}. 
In this regime,
$\mhh\sim\mha\sim\mhpm$, the $\hl$ has nearly SM-like properties, while
the $\hh$ will have weak $WW,ZZ$ couplings. Consequently,
the $\hl$ is the most experimentally accessible Higgs boson of the MSSM.
At the Tevatron (see \cite{tevreport}), 
integrated luminosity of order $15~\fbi$ ($20~\fbi$) is 
required to detect the $\hl$ if $\mha\lsim 250\gev$ ($\mha\lsim 400\gev$).
For $\mha=150\gev$ ($200\gev$) and $L=15~\fbi$, 
the $\ha,\hh$ will be detected in $b\anti b \hh+b\anti b\ha$
production if $\tanb>35$ ($\tanb>50$). 
The LHC (see \cite{LHCcms,LHCatlas} for the CMS and ATLAS
studies) is guaranteed to find one of the MSSM Higgs bosons with
$L=300~\fbi$ (roughly three years of high-luminosity operation),
but there is a significant wedge of moderate $\tanb$ where only
the $\hl$ will be detected unless SUSY decays
of the heavier Higgs bosons have substantial branching fraction.  
This is illustrated by the ATLAS plot of Fig.~\ref{atlasmaxmix}. 
Similar results
have been obtained by the CMS collaboration~\cite{LHCcms,cmsmssmhref}.

At a LC, the $\hl$ will be detected using the same production/decay modes
as for a light $\hsm$. In particular, the $\epem\to Z\hl$ and $\epem \to
\nu\anti\nu \hl$ (Higgstrahlung and $WW$ fusion) processes will
yield tens of thousands of $\hl$'s per year. However,
the $\hh,\ha,\hpm$ might not 
be detected in $\epem$ collisions at the LC~\cite{Grzadkowski:2000wj}.
First, the $\epem\to\hh\ha$, $\epem\to \hp\hm$, $\epem\to Z\ha\ha$
and $\epem\to \nu\anti\nu \ha\ha$ production
mechanisms would be forbidden for $\mha\gsim \rts/2$.
Second, while for very high (low) $\tanb$ it will be possible
to detect  $\epem\to b\anti b \ha,b\anti b \hh,tb\hpm$ 
($\epem\to t\anti t\ha,t\anti t \hh, tb\hpm$)
if not too near the relevant kinematic threshold,
there will be a wedge region of moderate $\tanb$
in which $t\anti t\hh+t\anti t\ha$
 and $b\anti b\hh+b\anti b\ha$ both produce too few events for detection. 
Assuming a substantial increase
in the LC $\rts$ is many years in the future, implementation of the 
$\gam C$ would be called for in order to
make possible the direct observation of the $\hh,\ha$
(through $\gam\gam\to \hh+\gam\gam\to \ha$)~\cite{Gunion:1993ce,gunasner,
Asner:2002aa,Muhlleitner:2001kw} for $\mha$ up to $\sim 0.8\rts$. 
A $\mu C$ with the same energy as the LC
would be able to detect the $\hh,\ha$ via the $s$-channel 
$\mu^+\mu^-\to \hh,\ha\to b\anti b,\tau^+\tau^-,t\anti t$
resonance signals using an appropriately designed energy scan 
procedure~\cite{Barger:1996jm,jfgucla,Barger:2001mi}.

\begin{figure}[t!]
\begin{center}
\vspace*{-2.5in}
\hspace*{.1in}\includegraphics[width=6in]{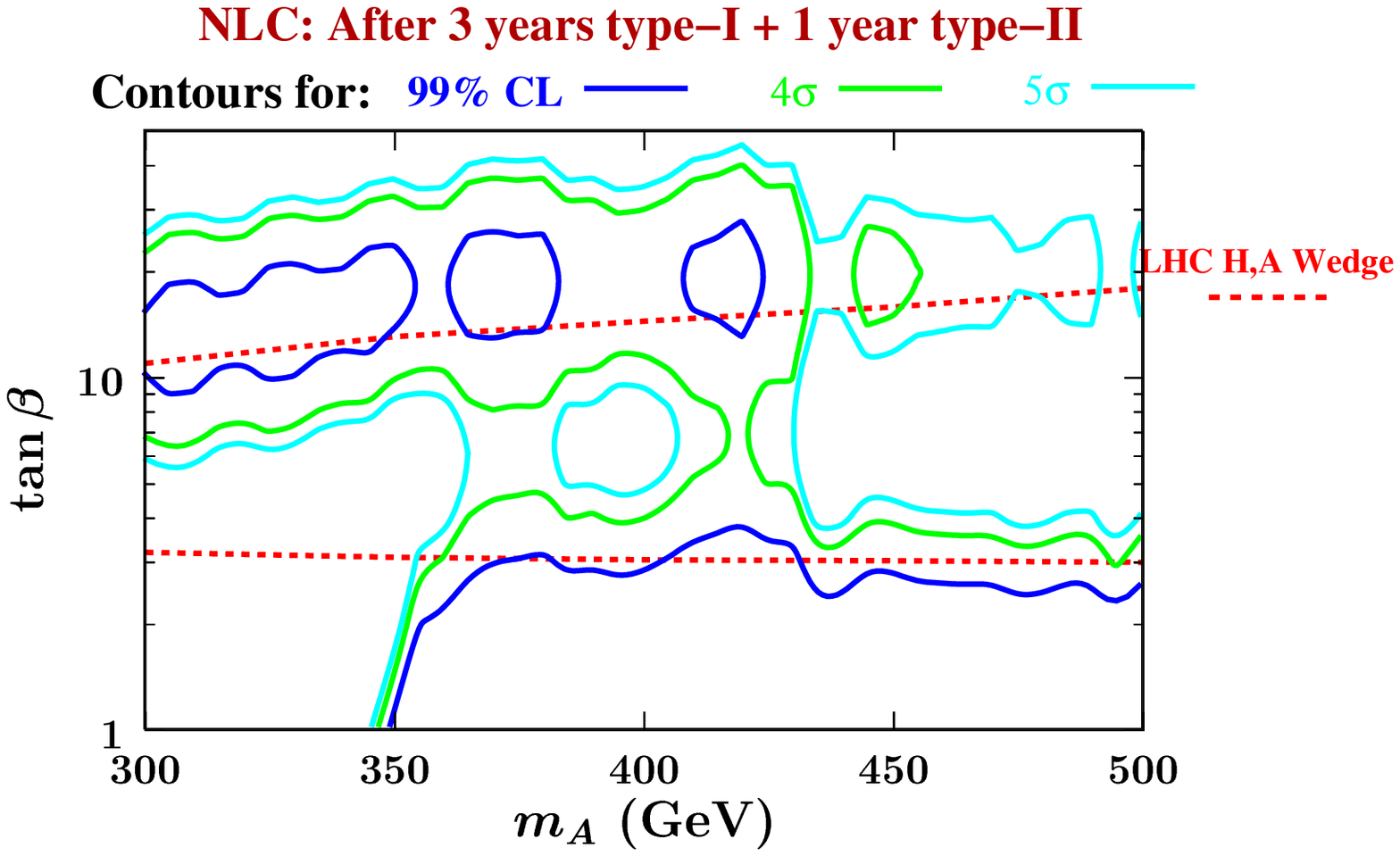}
\end{center}
\vspace*{-2.5in}
\capt{\label{NLCgamgam}
Regions for $\hh,\ha$ discovery and exclusion at a $\rts=630\gev$
LC based on the NLC/Livermore design, assuming 3 years of operation
in type-I configuration and 1 year in type-II configuration.
Higher luminosity, as possibly achievable at
TESLA, would be helpful to ensure coverage. Results are from \cite{gunasner,Asner:2002aa}.}
\end{figure}

In considering the $\gam C$ option, there are two distinct scenarios.
If precision $\hl$ measurements give a first indication
of the presence of the $\ha,\hh$ and a rough determination of $\mha\sim \mhh$
(both of which require knowing other MSSM parameters sufficiently
well to determine the size of the one-loop
corrections~\cite{Carena:2001bg} to the $b\anti b\hl$ coupling
and the extent to which premature or ``exact'' decoupling~\cite{Gunion:2002zf}
is present), then the $\gam C$ could be
set up to yield a $\gam\gam$ luminosity spectrum peaked
in the region of the expected $\mha\sim \mhh$ value.
Less than one year's
luminosity is needed for direct detection 
if you know $\mha$ within $\sim 50\gev$ (so that only two or three
different settings of $\rts$ are needed to explore the 
interval)~\cite{gunasner}.
However, if there is no indirect $\mha$ determination, or if there
is reason to mistrust the indirect determination (not an easy thing
to assess because of the possibility of large
corrections to the $b\anti b\hl$ coupling and/or premature
decoupling), the preferred approach would be to operate at the
highest $\rts$ available using several different $\gam C$
configurations. To illustrate, we summarize
the results of \cite{gunasner}, as reanalyzed in \cite{Asner:2002aa}, 
where it is supposed that the LC has $\rts=630\gev$.
Direct $\epem\to\hh\ha$ production is assumed to 
exclude $\mha\sim\mhh\lsim 300\gev$. 
The largest $\gam\gam$ energy for which good $\gam\gam$
luminosity could be achieved is $E_{\gam\gam}\sim 0.8\rts\sim 500\gev$.
To search in the full range between $300$ and $500\gev$, the optimal
approach is to employ two different configurations ({\bf I} and {\bf II},
as defined in Ref.~\cite{gunasner}) for the electron helicity / laser-photon
polarizations.
 The {\bf Type-II} configuration yields a $E_{\gam\gam}$ luminosity
spectrum peaked at the high end and would be used
to search the $\mha\sim\mhh\in[450,500]\gev$ interval.
The {\bf Type-I} configuration yields a
broader $E_{\gam\gam}$ spectrum with ability to probe
a range of lower masses, $\mha\sim\mhh\in[300,450]\gev$.
Both spectra types have substantial $\vev{\lam\lam'}$
of the back-scattered photons in the indicated mass regions, 
as needed to suppress the $\gam\gam\to b\anti b$ background
to the $\gam\gam\to \hh,\ha\to b\anti b$ signal.
Using this approach, Fig.~\ref{NLCgamgam} 
shows that a $\gam C$ based on the American/Asian
NLC design could detect the $\hh,\ha$ 
throughout most of the LHC wedge region at the $4\sigma$ level,
and exclude their presence at the 99\%CL throughout the entire
wedge, after about four years of operation.
Thus, if a light CP-even Higgs boson is detected at the LHC and LC,
but no heavier Higgs bosons, and if there are SUSY signals at the LHC
and LC consistent with moderate $\tanb$, a $\gam\gam$ collider
becomes mandatory in the absence of a timely upgrade
of the LC to higher $\rts$.

\smallskip
\noindent{\it Determining $\tanb$ in the MSSM using heavy Higgs bosons}
\smallskip

One of the most important parameters of the MSSM is $\tanb$.
While some measurements of $\tanb$ will be possible using gaugino
and slepton production, measurements of the Yukawa couplings
of the $\hh,\ha,\hpm$ provide the most direct measurement of the ratio
of vacuum expectation values that defines $\tanb$.
This is because in the decoupling regime the Yukawa couplings behave as
$t\anti t\hh,t\anti t\ha\propto \cotb$ and $\bb\hh,\bb\ha
\propto\tanb$. 

Simple observables sensitive to these couplings at a LC are:
a) the rate for $\bb\ha,\ \bb\hh \to \bb\bb$; 
b) the $\hh\ha\to \bbbb$ rate; 
c) a measurement of the average $\hh,\ha$ total width in $\hh\ha$ production;
d) the $\hp\hm\to \tbtb$ rate; and 
e) the total $\hpm$ width measured in $\hp\hm\to\tbtb$ production.
Because of limited experimental resolution for the width
measurements, the width determinations of $\tanb$ are only good
at high $\tanb$ where the intrinsic widths are large.
The rate determinations are typically only accurate at lower $\tanb$
values for which there is substantial variation of the $\hh,\ha\to \bb$
and $\hpm\to tb$ branching ratios.
If SUSY decays of the $\hh,\ha$ are present, this variation will
persist to higher $\tanb$ values.
The errors on $\tanb$ resulting from combining a)-e) above are
shown in Fig.~\ref{totalonly}, from \cite{ghjs}.

\begin{figure}[t!]
\begin{center}
\includegraphics[width=0.55\textwidth,angle=90]{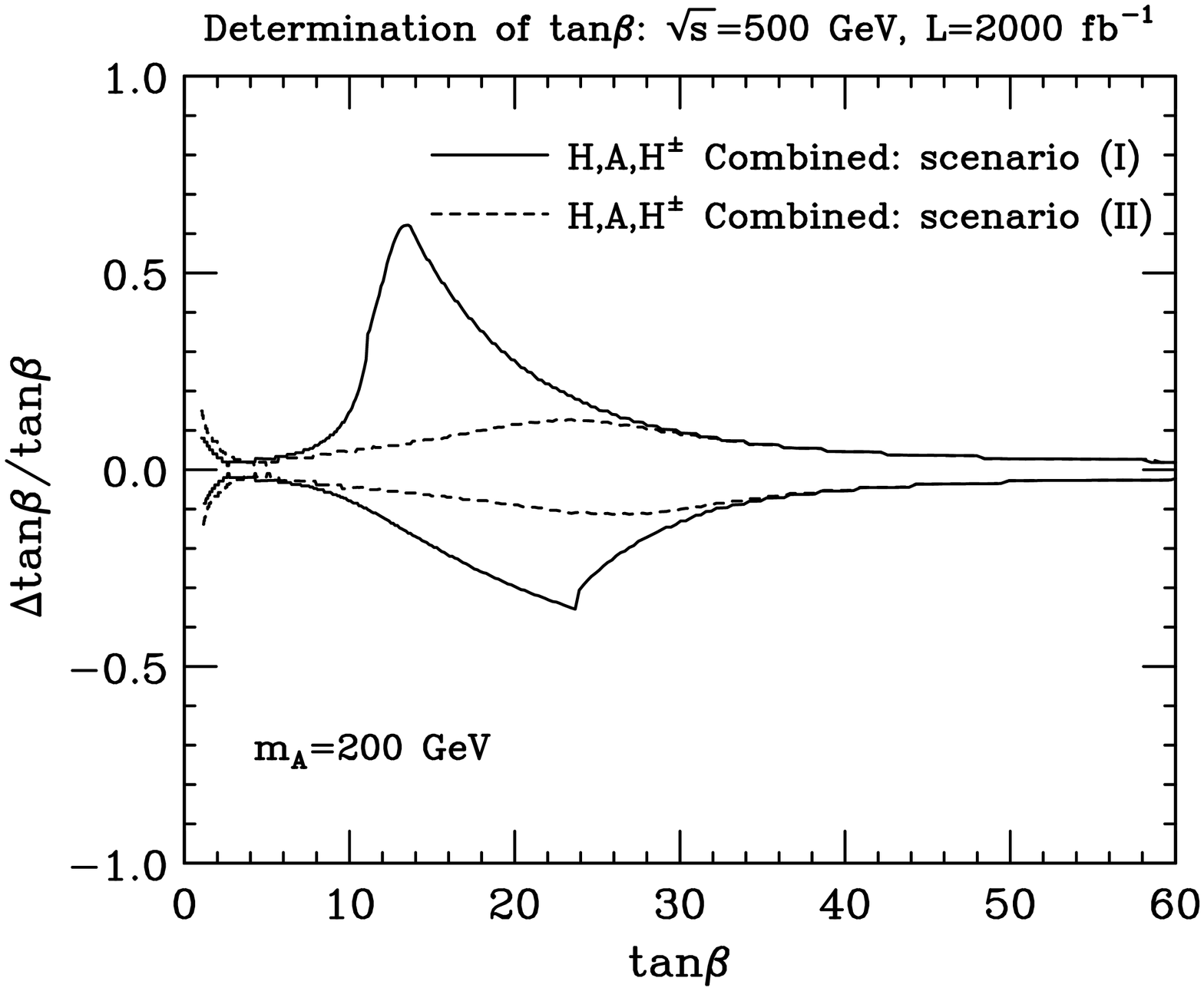}
\end{center}
\capt{\label{totalonly}
For the MSSM with $\mhpm\sim \mha = 200$~GeV, and 
assuming a LC with $L=2000~\fbi$ at $\rts=500\gev$,
we plot the $1\sigma$ statistical upper and lower bounds
in terms of $\Delta\tanb/\tanb$
as a function of $\tanb$ based on combining (in quadrature)
the results from the channels listed in the text.
Results are shown for two SUSY scenarios; in (I) SUSY decays
of the $\hh$ and $\ha$ are not present; in (II) $\hh,\ha\to\cnone\cnone$
decays are substantial. Results are taken from \cite{ghjs}.}
\vspace*{-.1in}
\end{figure}

We  note that
$\gam\gam\to \hh,\ha$ rates also provide a reasonably good $\tanb$
determination~\cite{gunasner} and would be the
only way of assessing the $\hh,\ha$ Yukawa coupling strengths
if the $[\mha,\tanb]$ parameter set lies in the wedge region.

Determining $\tanb$ at the LHC on the basis of heavy Higgs production
rates has been discussed
in \cite{Gunion:1996cn,LHCatlas,ghjs}. The LHC determination
may be superior in the $\tanb$ range from roughly 10 to 25
where the errors from the LC determination,
illustrated in Fig.~\ref{totalonly}, are largest. Most probably,
the width technique for determining $\tanb$ will not work at the LHC
except for really large $\tanb$ values.
This is because the $\hh,\ha\to\tauptaum$ 
channel (which is detectable in $\bb\hh+\bb\ha$ production
further down in $\tanb$ than other channels)
cannot be used for direct width reconstruction
because of the poor experimental width resolution, $\sim 15\%$.
Once $\tanb$ is very large,
detection of $b\anti b\hh+b\anti b\ha$ production with $\hh,\ha\to \bb$
will become possible, but even better,
the $\hh,\ha\to\mupmum$ decays will become visible and
provide an excellent intrinsic width measurement.  Detailed studies
have not been performed.

\smallskip
\vspace*{.2in}
\noindent{\it A CP-violating MSSM Higgs sector}
\smallskip

Generically, it is certainly possible that the soft-SUSY-breaking parameters
of the MSSM are complex.  If so, the one-loop
corrections to the Higgs tree-level potential can give rise to
CP violation in the Higgs sector. In this case, the MSSM Higgs
sector becomes rather similar to the CP-violating 2HDM, except
that there is still an upper bound on the mass of the lightest 
of the three neutral Higgs bosons, $h_{1,2,3}$. 
For this situation it is convenient to 
use $m_{h_1}$ (in place of $\mha$ for the CP conserving case),
$\tanb$  and a CP-violation angle $\phi$ to parameterize the
Higgs sector of the MSSM. 
A recent study~\cite{Carena:2002bb} examines a particular MSSM scenario
of this type, dubbed the CPX benchmark scenario, for
which CP violation in the Higgs sector can be substantial without
having electric dipole moments (EDM's) that violate current bounds.
As anticipated in our discussion of special cases in the general 2HDM, for 
large CP violation ($\phi=60^\circ$ or $90^\circ$) there are
portions of the $(\mhpm,\tanb)$ parameter plane where none of the Higgs
bosons of the MSSM can be detected at LEP~2, the Tevatron or the LHC.
In particular, one such region is characterized by $\phi=90^\circ$,
$m_{h_1}<60\gev$ and $\tanb\sim 3-5$.
At LEP~2, the $Zh_1$ Higgstrahlung signal is suppressed by weak
$ZZh_1$ coupling while the $h_2$ is either too heavy to be produced
or decays to $h_1h_1$, a signal for which existing LEP~2 analyses
are not well suited.  (This region might be excluded by a LEP~2
analysis focusing on 6-body final states.)
At the Tevatron and LHC none of the Higgs bosons are
detected by virtue of the fact that the heavier $h_{2,3}$ 
are the only Higgs bosons that have substantial couplings to
$WW,ZZ$, $t\anti t$ and $b\anti b$. But, despite abundant production rates they
cannot be detected because they decay to a pair
of lighter Higgs bosons or a lighter Higgs boson and the $Z$ 
(\eg\ $h_2\to h_1h_1,Zh_1$) --- 
the corresponding signals associated with the resulting
final states, such as $h_1h_1\to\bb\tau^+\tau^-$, have not been shown to
be observable in the presence of expected backgrounds.  
There are also cases in which the $h_{1,2,3}$ signals in
a given discovery channel are all of similar size and
overlap due to limited experimental resolution  --- 
there is no demonstration by the ATLAS and CMS collaborations
that the  resulting broad enhancement would be distinguished from
the background.

An LC with $\rts\sim 500\gev$ would be guaranteed to find at least
one of the $h_{1,2,3}$ since the model constrains the $h_i$
with the largest $g_{ZZh_i}^2$ coupling to be fairly light. If
it decays substantially to a still lighter Higgs boson pair, then
the latter could also be studied. On the other hand,
there is a distinct possibility that one of the three $h_i$
does not have highly enhanced $b\anti b$ coupling and 
does not appear in the decays of a heavier Higgs boson; to
detect it would probably require the $\gam C$ or a $\mu C$ --- see
the earlier discussion regarding the MSSM $\ha$~\cite{gunasner,jfgucla}.

\smallskip
\noindent{\it The Next to Minimal Supersymmetric Model, NMSSM}
\smallskip

Let us now turn to the NMSSM model in which one adds an extra singlet
superfield to the MSSM (see \cite{hhg} for a summary of the NMSSM).
This provides an extremely natural source for the $\mu$
term of the MSSM via the superpotential term
$W\ni \lam \what H_1 \what H_2 \what N$. When  
$\vev{(\what N)_{\mbox{scalar component}}}=n$, where $n$ of order
 the electroweak scale is natural in many cases,
an effective $\mu_{\rm eff}\sim \lam n$ results. 
(Note that $n$ can be traded for $\mu_{\rm eff}$ in describing parameter space.)
Another possible superpotential terms is $\kappa \what N^3$.
Assuming no CP violation, the NMSSM
Higgs sector will have an extra complex scalar field in
addition to the usual two doublet fields, resulting in
three CP-even Higgs bosons, $h_{1,2,3}$, two CP-odd Higgs bosons,
$a_{1,2}$, and a charged Higgs pair, $\hpm$.

Many groups have shown that a LC will find at least
one of the CP-even Higgs bosons of the NMSSM (\eg\
via the Higgstrahlung process) for any choice of $\lam$ and $\kappa$
consistent with perturbativity up to high scales.
A recent study appears in \cite{Ellwanger:1999ji}.
The keys are that the Higgs bosons must share
the net $VV$ coupling squared of the SM Higgs boson and that
the sum of the Higgs masses squared times their $VV$ couplings-squared
has a strong upper bound in the perturbative NMSSM context.
However, the situation at the LHC
is far more uncertain. At the time of Snowmass96, it was demonstrated
\cite{Gunion:1996fb} that one could 
find parameter choices for Higgs masses and mixings such
that the LHC would find no Higgs boson 
using just the production/detection modes explored
up to that time. Since then, there have been some improvements
in LHC simulations and new discovery channels have been added.
In \cite{Ellwanger:2001iw}, it was shown that Higgs discovery
for all of the difficult parameter choices identified in the Snowmass96
work would be possible in the newly analyzed $t\anti t\h\to t\anti t b\anti b$
mode \cite{6r,6.01r,6.02r}. 
Ref.~\cite{Ellwanger:2001iw} also 
shows that the addition of $WW$ fusion discovery modes (as studied
for the SM Higgs boson in \cite{6.2r})
will allow detection of at least one NMSSM Higgs boson for all
parameter choices, {\it  provided we exclude choices for which a heavier Higgs boson
decays primarily to a pair of lighter Higgs bosons}.

In more detail, the modes employed in 1996 were:
(1)~$gg\to \h\to\gam\gam$ at LHC; 
(2)~$W\h,t\anti t\h \to \ell+\gam\gam$ at LHC; 
(3)~$gg\to\h,\a\to\tauptaum$ plus 
$b\anti b \h,b\anti b\a\to b\anti b\tauptaum$ at LHC;
(4)~$gg\to \h\to ZZ^*~{\rm or}~ZZ\to 4\ell$ at LHC;
(5)~$gg \to \h\to WW^*~{\rm or}~WW\to 2\ell 2\nu$ at LHC;
(6)~$\zstar\to Z\h$ and $\zstar\to \h\a$ at LEP2. 
To these, \cite{Ellwanger:2001iw} added
(7)~$gg\to t\anti t\h\to t\anti t b\anti b$; 
(8)~$WW\to \h\to \tauptaum$;  and
(9)~$WW\to \h\to WW^{(*)}$. 
If one avoids regions of parameter space where
(a)~$\h\to\a\a$,
(b)~$\h\to\h'\h'$,
(c)~$\h\to\hp\hm$,
(d)~$\h\to \a Z$,
(e)~$\h\to \hp\wm$,
(f)~$\a\to\h\a'$,
(g)~$\a\to Z\h$, and 
(h)~$\a\to \hp\wm$
are present, and where  
(i)~$\a,\h\to t\anti t$, (j)~$t\to \hpm b$ decays are possible
then discovery of at least one NMSSM Higgs boson is always possible
at the LHC. The parameters varied comprised 
$\lam$, $\kappa$, $\mu_{\rm eff}$, $\tanb$, 
$A_\lam$, $A_\kappa$. Constraints from renormalization group evolution and
perturbativity were imposed. The most difficult points found for
the LHC have marginal rates for the following reasons.
First, the $WW,ZZ$ coupling-squared is shared among the $h_i$
($\sum_i g_{VVh_i}^2=g_{VV\hsm}^2$ is required). This decreases the
decays and production processes that rely on the $VVh_i$ coupling.
In particular, it is easy to make
the $\gam\gam$ coupling and decays small since the reduced $W$ loop cancels
strongly against $t,b$ loops. Second,
since $\tanb$ is not very large one is
well inside the `LHC wedge' (as discussed earlier) for all Higgs bosons.
As a result, one needs the
full $L=300~\fbi$ for ATLAS and CMS and the $WW$ fusion modes to achieve
an observable signal. In making the claim of observability here,
the partonic level
$WW$-fusion results of \cite{6.2r} were employed. 
These channels are still being studied by the LHC collaborations.

Of course, there is much more work to do on how to detect Higgs bosons
in Higgs pair or $Z$+Higgs decay modes at the LHC. 
The parton-level study of~\cite{Dai:1995cb}
suggested that in the MSSM the $\hh\to\ha\ha \to 4b$ 
process could be detected by
using 3 or 4 $b$ tagging, reconstructing the
double $\ha$ mass peak, and reconstructing the $\hh$ mass peak.
Studies by the LHC experimental collaborations are casting doubt
that this signal will actually be observable \cite{moretti}.
In any case, the MSSM 
results also need to be translated into the NMSSM context.
The $WW\to h_i\to a_ja_j, h_kh_k$ modes could also prove
extremely valuable, but have not yet been simulated.

\smallskip
\noindent{\it A continuum of Higgs resonances}
\smallskip

One of the most difficult cases~\cite{Espinosa:1998xj}
for Higgs discovery is 
when there is a series of Higgs bosons separated by the mass
resolution in the discovery channel(s) --- \eg\ in $\epem\to Z+$Higgs
there would be one Higgs boson every $\sim 10\gev$ 
(the detector resolution in the recoil mass spectrum).
Since extra singlet and doublet representations
(beyond the minimal two-doublets required in SUSY models)
are abundant in string models,
this possibility deserves serious consideration.
In general, all the extra neutral Higgs bosons would 
mix with the normal SM Higgs (or the MSSM scalar Higgs bosons) in such
a way that the physical Higgs bosons share the $WW/ZZ$ coupling and
decay to a variety of channels. The only iron-clad approach would then
be to use $\epem\to Z+X$ production and look for a broad excess in 
the recoil mass, $\mx$.
Fortunately, there are significant constraints on this scenario.
Adopting a continuum notation, we have
\beq
\int_0^\infty 
dm K(m)m^2=m_C^2\,, \quad\mbox{where}\quad \int_0^\infty K(m)=1
\label{msqlim}
\eeq
where $K(m)(gm_W)^2$ is the (density in Higgs mass of the) 
strength of the $hWW$ coupling-squared.
Precision electroweak data suggests $m_C^2\lsim (200-250\gev)^2$
in the absence of compensating $\Delta T>0$ contributions from
some heavy Higgs bosons or other new physics.
Further, for multiple Higgs representations 
of any kind in the most general SUSY context, the RGE equations plus 
perturbativity {up to $\mgut\sim 2\times 10^{16}\gev$}
gives the same constraint on $m_C$.
An OPAL analysis \cite{Abbiendi:2002qp} of their LEP2 data
in a decay mode independent fashion imposes strong
constraints on possible weight $K(m)$ in the region $K(m)<80\gev$.
In particular, using data for $\epem\to ZX$
with $Z\to \epem$ or $\mupmum$, they obtain
an upper limit (at 95\% CL) on $\int_{m_A}^{m_B} dm K(m)$ for any
choice of $m_A$ and $m_B$. 
For example, for $m_A=0$, they have eliminated almost the full interval up to
$m_B\sim 350\gev$ assuming $m_C=200\gev$.
But, for $m_A\geq 80\gev$, they have not eliminated any interval.


To go further, requires higher energy.  A LC energy of
$\rts=500\gev$ is more or less ideal. The required analysis
is given in  \cite{Espinosa:1998xj}.
If we assume that $K(m)$ is constant, that
$m_C=200\gev$, and that $m_A=70\gev$, then $m_B=300\gev$.
A fraction $f=100\gev/230\gev \sim 0.43$ of the continuum Higgs signal 
then lies in the $100-200\gev$ region. (This interval is chosen to avoid 
the $Z$ peak region with largest background while avoiding
kinematic suppression of
the $Z+$Higgs cross section when $\rts=500\gev$.)
Summing $Z\to\epem+\mupmum$ leads to an integrated
signal rate of $S\sim 540 f$ with a background
rate of $B=1080$  for the  $100-200\gev$ window, assuming $L=200~\fbi$.
The result is
$
{S\over\sqrt B}\sim 16f \left({L\over 200~\fbi}\right)\mbox{ for } m\in[100-200]\gev\,$.
This is a robust signal that would be easily detected.
With $L=500~\fbi$, one can determine the magnitude of the signal
with reasonable error ($\sim 15\%$) in
each $10\gev$ interval of $\mx$.  This is a clear case in which
the LC would be essential for observing and studying Higgs bosons
since detection of this kind of continuum signal at a hadron collider
appears to be almost certainly impossible.

\smallskip
\noindent{\it Left-right symmetric supersymmetric models}
\smallskip

Finally, let us consider the left-right symmetric supersymmetric model
(LRSSM)~\cite{Mohapatra:1996vg,Mohapatra:1997su,Babu:2001se,Babu:2002tb}.
In general, the LR-symmetric models assume that nature has an underlying parity
invariance and it is Higgs fields that break the parity at some high scale.
The group structure prior to breaking is typically taken to be 
SU(2)$_L\times$SU(2)$_R\times$U(1)$_{B-L}\times$SU(3)$_C$
and it is the SU(2)$_R\times$U(1)$_{B-L}$ symmetry that is broken 
down to U(1)$_Y$ at scale $m_R$. 
The above
groups are naturally contained within SO(10), the fundamental representations
of which automatically contain $\nu_R$ fields as well as the 
SU(5) representations for the observed fermions.
Higgs fields are easily introduced in such a way that a large Majorana mass
is induced for the $\nu_R$ when parity 
is broken, leading to the see-saw mechanism for
neutrino masses. Further,
the LRSSM context guarantees that,
at scale $m_R$, there is no strong CP problem and 
no SUSY-CP problem (\ie\ the generic problem of SUSY phases
for the $\mu$ parameter and for the gluino mass that would yield
large EDM's unless cancellations are carefully arranged).
It is then a matter of making sure that evolution from $m_R$ down
to the TeV scale does not destroy these latter two properties.

In fact, there are two LRSSM's on the market.  In one, there is Majorana
lepton-flavor-violation (LFV) as referred to above while in the other
the LFV is Dirac in nature. In the former, the superpotential includes
the generic terms (I will drop the $\what{}$ notation for superfields) 
$W\ni f \nu^c\nu^c \Delta+Y_\nu\nu^c  L
 H^u$, where the fermionic component of $ \nu^c$
is the $\nu_R$, $\Delta$ is a $B-L=2$ Higgs superfield 
the scalar component of which is the $\Delta_R$ Higgs triplet
representation discussed earlier, and  $ H^u$ is the superfield whose
scalar component is the Higgs doublet 
field, with neutral component vacuum expectation value $v_L$,
responsible for up-type quark masses.
For $\vev{\hzero_R}\sim m_R$, one generates the required Majorana $\nu_R$
mass and at scales $\ll v_R$ the $\nu_L$ masses
will be of order $Y_\nu^2v_L^2/m_R$, \ie\ very small. 
In the Dirac LFV models, the mass matrix containing $m_R$ is either
put in ``by hand'' as a bare mass terms in the Lagrangian (such terms
are super renormalizable) or arises from non-renormalizable operators involving
a $B-L=1$ Higgs boson $\chi^c$ via couplings of the type
$( \nu^c \chi^c)^2/M$.  Thus,
Majorana neutrino mass generation is rather ad hoc in the Dirac LFV
models. Further, the Dirac LFV models are most
attractive for a large $m_R\sim M_U$ scale (which allows for MSSM-like
coupling constant unification). This makes
the model of less interest for TeV scale experimentation.  Thus, I will
focus on the Majorana LFV LRSSM case, in which the scale $m_R$
is (given current theoretical results) required to be of order a TeV,
in which case all the exotic Higgs bosons would be potential accessible.

The matter ({\it i.e.} not related to the gauge bosons of the model) 
superfields required in the Majorana LFV LRSSM model are as follows.
(Here, I will drop the $L,R$ subscripts and use the notation
$\Delta\equiv \Delta_L$, $\Delta^c\equiv \Delta_R$
of the references given earlier.)  
(a) Two bi-doublets $\Phi_{1,2}$, with 
SU(2)$_L\times$SU(2)$_R\times$U(1)$_{B-L}$ quantum numbers ($2,2,0$), 
are required in order that the vacuum expectation values
of the neutral spin-0 components lead to a CKM matrix
that is not simply proportional to the identity matrix.
(b) An SU(2)$_R$ triplet $\Delta^c$ ($1,3,+2$) is required, whose
neutral spin-0 component breaks SU(2)$_R$ symmetry when it acquires a
vev of order $m_R$.
(c) The corresponding 
SU(2)$_L$ triplet $\Delta$ ($3,1,+2$) is required by L-R symmetry.
(d) In addition, the $\Delta$ and $\Delta^c$ must have anti-field partners,
$\anti\Delta^c$ ($1,3,-2$) and $\anti\Delta$ ($3,1,-2$)
in order that all anomalies cancel. 
(e) We also require the quark and lepton superfields,
$ Q$  $(2,1,1/3)$,
$ Q^c$  $(1,2,-1/3)$,
$ L$  $(2,1,-1)$, and 
$ L^c$  $(1,2,+1)$,
whose spin-1/2 components are the normal quarks and leptons.
(f) Finally, there may be a CP-odd singlet $(1,1,0)$ superfield
that breaks the parity symmetry when its scalar component
acquires a vev and a CP-even partner singlet. (The alternative 
for achieving breaking of P is certain nonrenormalizable interactions.)

We give a very brief summary of
how the LRSSM models avoid the strong CP
and SUSY CP problems. This is accomplished as follows.
Consider first the strong CP problem.
The standard strong CP quantity is
\beq
\anti \Theta=\Theta+{\rm Arg}~{\rm det}(M_uM_d)-3{\rm Arg}(m_{\wtil g})
\eeq
where $\Theta$ is the coefficient of the $F_{\mu\nu}\wtil F^{\mu\nu}$
term (which is P violating)
and $\anti \Theta$ must be very small to solve the strong CP problem.
The P invariance for scales above $m_R$ guarantees that
$\Theta=0$ above $m_R$.
The L-R symmetry requires that $m_{\wtil g}$ be real above $m_R$.
Finally, the
Yukawa coupling matrices are required to be hermitian by the L-R symmetry
 transformations. Then, if the bi-doublet Higgs vevs 
are real the quark mass matrices
will be hermitian, which in turns implies that the determinant 
of the 2nd term is real.
Note that it is necessary to show
that the required Higgs potential does not
give rise to spontaneous CP violation.  It turns out that
this is not really automatic~\cite{Mohapatra:1997su}; 
problematical phases develop at one loop unless the scale $m_R$ is
of order $\msusy\sim \tev$.
The other possibly unnatural feature of the Majorana LFV approach
is that a single non-renormalizable operator
${\lam\over M}[{\rm Tr}(\Delta^c\tau_m\anti\Delta^c)]^2$ 
($M$ is of order $\mpl$ or $m_R$) is needed 
in order that the  vacuum state of the model have
$\vev{\snu_R}=0$ (so that R-parity is conserved).

Regarding the SUSY CP issue, we first note
that, generically speaking, it is necessary
to have small phases for $Am_{\wtil g}$ and $\mu v_u m_{\wtil g}/v_d$.
At scales above $m_R$, the hermiticity of $A_u$ and $A_d$ 
(the soft-SUSY-breaking parameter matrices) and
of the Yukawa coupling matrices, along with reality of $m_{\wtil g}$,
guarantees the required reality. 

The result is that the naturalness, strong CP and SUSY CP
problems can all be solved in the context of the LRSSM
without R-parity violation.  Further, in the Majorana LFV case this 
requires many Higgs fields, including
SU(2)$_L$ and SU(2)$_R$ triplets as well as doublets, and a low
scale for $m_R$ that would imply TeV scale masses for all these
new Higgs bosons (as well as for the $W_R$).
Measuring their properties would be 
key to understanding the full structure of the LRSSM model.
The two downsides of having $m_R$ of order a few TeV are: (i)~generating 
small neutrino masses via the see-saw mechanism requires
careful adjustment, \ie\ small values,
of the associated lepton-number violating couplings; and (ii)~coupling
unification is hard to arrange and would typically require
extra matter and/or extra dimensions.

\vspace*{-.1in}
\section{\bfbm Determining the CP nature of a Higgs boson}

In essentially all of the extended Higgs scenarios considered above,
either there are one or more CP-odd Higgs bosons (for
the case of a parity conserving Higgs sector) or a collection of
Higgs bosons of mixed CP nature.  Direct determination of the CP nature
of any observed Higgs boson
will probably be critical to disentangling any but 
the simplest SM Higgs sector. For this the $\gam C$ facility would be 
ideal~\cite{Grzadkowski:1992sa,Gunion:1994wy,Kramer:1993jn,gunasner}.
A muon collider would also be of great value for
determining the CP nature of observed Higgs bosons~\cite{Barger:1996jm,Grzadkowski:2000hm,Atwood:1995uc}.

Let us focus on the $\gam C$.
We recall that the $\sig(\gam\gam\to \hh)\propto \vec\eps_1\cdot
\vec \eps_2$ while $\sig(\gam\gam\to\ha)\propto \vec\eps_1\times\vec\eps_2$.
Thus, if you could produce 100\% transversely polarized
back scattered photons, only the $\ha$ ($\hh$) would be produced
for perpendicular (parallel) polarizations, respectively.
In practice, there is always some circular polarization for the
back-scattered photons, even for 100\% transversely polarized laser
photons. Also, it could be that the Higgs bosons are of mixed CP parity.
(Although, in the decoupling limit the light Higgs boson is guaranteed
to be CP-even~\cite{ghk}.)
Thus, to fully explore the CP parity of a Higgs boson,
measurements of three asymmetries, $\cala_{1,2,3}$ would be ideal.
These are defined as
\bea
&\cala_1={|\calm_{++}|^2-|\calm_{--}|^2\over |\calm_{++}|^2+|\calm_{--}|^2 }
\,,\quad 
\cala_2={2{\rm Im}\left(\calm_{++}\calm_{--}^*\right)\over |\calm_{++}|^2+|\calm_{--}|^2 }\,,&\nn\\
&\cala_3={2{\rm Re}\left(\calm_{++}\calm_{--}^*\right)\over |\calm_{++}|^2+|\calm_{--}|^2 }={|\calm_{\parallel}|^2-|\calm_{\perp}|^2\over|\calm_{\parallel}|^2+|\calm_{\perp}|^2 }\,.&
\eea
The first two asymmetries are typically quite substantial 
for a large range of 2HDM parameter space for which CP
violation occurs. $\cala_3=+ 1$ ($-1$) for a purely CP-even (CP-odd)
Higgs boson.
In terms of the Stokes parameters specifying the polarizations
of the back-scattered photons
\bea
dN&=&dL_{\gam\gam}dPS\quarter \left(|\calm_{++}|^2+|\calm_{--}|^2\right)\times\nn\\
&&\quad
\left[ (1+\vev{\xi_2\xi_2^\prime})+(\vev{\xi_2}+\vev{\xi_2^\prime})\cala_1
+\left(\vev{\xi_3\xi_1^\prime}+\vev{\xi_1\xi_3^\prime}\right) \cala_2
+\left(\vev{\xi_3\xi_3^\prime}-\vev{\xi_1\xi_1^\prime}\right)\cala_3\right]\,.
\eea
The actually measured asymmetries are then
\bea
&T_1={N_{++}-N_{--}\over N_{++}+N_{--}}={\vev{\xi_2}+\vev{\xi_2^\prime}
\over 1+\vev{\xi_2\xi_2^\prime} } \cala_1\,, \quad
T_2={N(\phi={\pi\over 4})-N(\phi=-{\pi\over 4}) \over 
N(\phi={\pi\over 4})+N(\phi=-{\pi\over 4})}
={\vev{\xi_3\xi_1^\prime}+\vev{\xi_1\xi_3^\prime}
\over 1+\vev{\xi_2\xi_2^\prime}}\cala_2\,,&\nn\\
&T_3={N(\phi={\pi\over 2})-N(\phi=0) \over 
N(\phi={\pi\over 2})+N(\phi=0)}
={\vev{\xi_3\xi_3^\prime}-\vev{\xi_1\xi_1^\prime}
\over 1+\vev{\xi_2\xi_2^\prime}}\cala_3\,,&
\eea
where for $T_1$ we 100\% polarize the laser photons both with $+$
helicity and then flip both to negative helicities and
for $T_{2,3}$ $\phi$ is the angle between the 100\%
linear polarizations of the laser photons.
$T_2$ and $T_3$ are harder to measure than $T_1$
because the Stoke's parameters
in the numerators  are smaller for the former two. Nonetheless, excellent
accuracy can be achieved.
For example, at the American/Asian NLC
with the LLNL laser and IP design,
$\cala_3$ can be measured to $10\%$ after two years
of dedicated operation in the case
of a $120\gev$ CP-even SM-like Higgs boson~\cite{gunasner}. 
Similar accuracy can be achieved at the $\mu C$~\cite{Grzadkowski:2000hm}
using asymmetries~\cite{Barger:1996jm,Atwood:1995uc} 
obtained by varying the polarizations of the
colliding $\mu^+$ and $\mu^-$ in $\mu^+\mu^-$ $s$-channel Higgs production.

This accuracy should be compared to what is possible at the LHC and
the LC. At the LHC, parton level studies~\cite{Gunion:1998hm} suggest
some promise for determining the relative size of the CP-even and CP-odd
couplings of a Higgs boson to $t\anti t$ by looking at angular
distributions of the $t$, $\anti t$ and Higgs boson relative to one another,
where the Higgs boson is detected in the $\gam\gam$ (for a SM-like Higgs boson)
or $b\anti b$ (for other types of Higgs bosons) decay mode.  More realistic
studies are only now being performed but show less promise.

There are numerous studies of Higgs CP-determination using $\epem$
collisions (see, \eg, \cite{Kramer:1993jn} and \cite{Schumacher:2001ax}).  
However, caution is necessary in interpreting the
results of those that rely on angular distributions and the
like in the $Z+$Higgs final state.  Using $h$ ($a$) to denote
a CP-even (CP-odd) canonically normalized state,
a mixed-CP Higgs state can be written in the form
 $h_M=\cos\phi_M h+\sin\phi_M a$. The crucial point is that
the $aZZ$ coupling is at the one-loop 
level~\cite{Gunion:1991cw} compared to the tree-level $hZZ$
coupling. The cross section 
$d\sig/d\cos\theta$ for the $h_M$ contains terms 
proportional to ($L$ is a typical one-loop factor)
$\cos^2\phi_M$ and $\sin^2\phi_M L^2$ that are even in $\cos\theta$
and a term proportional 
to $\cos\phi_M\sin\phi_M L$ that is odd in $\cos\theta$,
and provides the best sensitivity to the $a$ component. 
For the $a$ component to have a strong fractional influence, requires 
$\tan\phi_M L\sim 1$.
In this case, 
all the terms in $d\sig/d\cos\theta$
will be of order $L^2$, including the term odd in $\cos\theta$,
and errors for the CP determination will be very large since the 
$h_M$ production rate will be small.
If $\cos\phi_M$ is substantial, the rate will be large
but the fractional influence of
the $a$ component will be at the one-loop level and very hard to detect.
 This same
caution applies to CP determinations related to $h_M\to \wp\wm$ or $ZZ$
decay angular distributions (see, \eg, \cite{Soni:jc,Skjold:1994qn,Han:2000mi}).

The best technique for the $Zh_M$ final state is to employ 
the self-analyzing decays $h_M\to \tau^+\tau^-$. We wish to probe
the relative strengths of the $1$ versus $\gamma_5$ terms
in the interaction $\anti\psi_\tau(\cos\phi_M+i\gam_5\sin\phi_M)\psi_\tau$.
In the simple limits of $\cos\phi_M=1,0$, respectively, one finds
$
\Gamma(h_M\to\tau^+\tau^-)
\propto (1-s_{\parallel}^{\tau^+}s_{\parallel}^{\tau^-}\pm s_\perp^{\tau^+}s_\perp^{\tau^-})\,,
$  
where $\parallel,\perp$ denote components parallel/transverse 
to the Higgs boson
momentum as seen from the respective $\tau^\pm$ rest frames.
(The corresponding expression in
the general case is complicated.)
While these spin directions are not directly measurable, the 
distributions of the $\pi^\pm$ or $\rho^\pm$ from the $\tau^\pm\to \pi^\pm\nu$
or $\tau^\pm\to \rho^\pm\nu$ decays will reflect the
the spin directions and one can extract the relative magnitude
of the CP-even versus CP-odd coupling. This technique
shows substantial promise according to theoretical 
studies~\cite{Kramer:1993jn,Grzadkowski:1995rx}. A more detailed
experimental study~\cite{Bower:2002zx},
using somewhat different techniques than
originally proposed, finds that the CP-even
nature of a $\h$ with $\mh=120\gev$ can be verified at
the 95\% CL in $Z\h$ production at $\rts=500\gev$, assuming $L=500~\fbi$.
Thus, for a CP-even $\h$ the $\gam C$ initial state polarization
asymmetries and the final state LC $\tau^+\tau^-$ analysis 
yield comparable accuracies.  However, since the $aZZ$ coupling
is one-loop, 
$\epem\to Za$ production will have low rate and
only the $\gam C$ (or $\mu C$) could verify the CP nature
of a state that is mainly or entirely CP-odd. It should also be
noted that if $\tau^+\tau^-$ decays are suppressed (\eg\ because
of competing Higgs pair final states and/or SUSY final states),
the accuracy of the $\tau^+\tau^-$ technique will suffer, whereas
the $\gam C$ (and $\mu C$) asymmetry measurements are for production rates,
and are independent of how the Higgs boson decays. Finally, we note
that the $\tau^+\tau^-$ final state CP determinations 
performed for a Higgs produced at a $\gam C$ 
(or $\mu C$~\cite{Grzadkowski:1995rx,Atwood:1995uc})
would complement the determination
obtained using initial state polarization asymmetries.

\vspace*{-.1in}
\section{\bfbm Conclusions}

There are many quite well-motivated possibilities for
the Higgs sector that go far beyond the one-doublet sector of the 
SM. The plethora of possibilities means that it is entirely possible
that the Higgs sector will prove very challenging to fully explore.
The variety of models, complications
due to unexpected decay modes (\eg\ Higgs pairs or SUSY particles), 
overlapping of resonances,
sharing of $WW,ZZ$ coupling strength, CP violation, 
the possible impact of extra dimensions and Higgs-radion mixing, \etc\ 
make attention to multi-channel, multi-collider analysis vital.
In particular, it seems we must accept the fact that
there is enough freedom in the Higgs sector that we
should not take Higgs discovery at the Tevatron or LHC for granted
and that even at the LC Higgs detection and study 
could prove quite challenging (as in the light-$\ha$ scenario
for the general 2HDM where $\mhl$ can be as heavy as $\sim 800-900\gev$
without conflicting with precision electroweak data or perturbative
constraints).
The LHC collaborations must keep improving and 
working on every possible signature and the LC design must
be pushed to the highest feasible energy given financial and technological
constraints. Research regarding the feasibility of a $\mu C$ should
be continued.

The LHC ability to show that the $WW$ sector is perturbative could be
very useful.  Two particular examples are the following.
First, in the NMSSM we might not detect a Higgs
boson using the analysis techniques considered so far, 
but a perturbative $WW$ sector would imply that 
there are light CP-even Higgs bosons
with significant $WW$ coupling. Perhaps with that motivation,
it would be possible to find new techniques capable
of digging out faint signatures.
Second, we can imagine a scenario
in which there are a number of heavy $\sim 800-900\gev$ mixed-CP Higgs 
bosons~\footnote{Current precision electroweak constraints 
would be satisfied due to
weak-isospin breaking arising from mass differences relative
to charged Higgs bosons.}
that share the $WW,ZZ$ coupling strength strength and/or they
decay to lighter Higgs bosons (with small $ZZ$ coupling) and/or they give rise
to overlapping resonance signals. In such a scenario, it would
be impossible to absolutely guarantee discovery
of a Higgs boson at the LHC or at the  LC, $\gam C$ or $\mu C$
unless the center of mass energy of the latter machines
can reach the multi-TeV level. 
At the LHC, the $WW$ scattering processes would 
exhibit moderately perturbative 
behavior, and Giga-$Z$ operation at the LC
would show that the $S,T$ values matched the expectations for
such a scenario. These observations would indicate the need
for sufficiently higher $\rts$ at the LC to make production
of a pair of the CP-mixed Higgs bosons possible.

Sticking to less extreme and better-motivated cases
in which one or more Higgs bosons are reasonably light, 
it seems very apparent that experimentation at both the LC and 
the LHC is needed to have a high probability of discovering 
even one Higgs boson and almost certainly 
both machines will be needed to fully study the Higgs sector.
Particularly strong motivations for the LC, $\gam C$ and $\mu C$
include the following.
The LC would possibly be necessary in the case of the NMSSM 
and would certainly be required to detect a continuum
of strongly mixed CP-even Higgs bosons.
Observation of the heavy $\hh,\ha$ of the MSSM 
will require $\gam\gam$ collisions
if $[\mha,\tanb]$ are in the ``wedge'' region of parameter space.
Once observed, the properties and rates
for the $\hh,\ha$  will help enormously in determining important SUSY
parameters, especially checking for the predicted relation
between their Yukawa couplings and $\tanb$.
Exotic Higgs representations, \eg\ the triplet as motivated
by the seesaw approach to neutrino masses and the LRSSM
solutions to the strong and SUSY CP problems, will lead to exotic
collider signals and possibilities that might ultimately
be best explored via $e^-e^-$ and/or $\mu^-\mu^-$ collisions.
Finally, we have reviewed how important a $\gam C$ (and eventual
$\mu C$) could be for directly measuring the CP composition
of a Higgs boson, especially one with a substantial CP-odd component.

In short, since our ability to fully explore the Higgs sector
will be very important to a full understanding of the ultimate theory,
it seems very clear that a full complement of collider facilities will
ultimately be needed, including the LHC, a LC, a $\gam C$
at the LC, and eventually a $\mu C$.

\centerline{\bf Acknowledgments}
\bigskip

This work was supported in part by the U.S. Department of Energy
and the Davis Institute for High Energy Physics.
I wish to thank P. Zerwas and DESY for their kind hospitality.
Of my many collaborators, I would like to especially
thank D. Dominici, B. Grzadkowski and H. Haber for
their many contributions.  


\end{document}